\documentclass{aa}  

\usepackage{hyperref}
\hypersetup{
    colorlinks=true,
    citecolor=blue,
    linkcolor=blue,
    urlcolor=blue,
}
\usepackage{xcolor}
\usepackage{natbib}
\usepackage{graphicx}
\usepackage{txfonts}
\usepackage{threeparttablex}
\usepackage{longtable}
\usepackage{booktabs}

\def\swift{{\it Swift~}}

\def\snxv{${\sigma^2_{NXV}}$}

\def\mbh{${\rm M}_{\rm BH}$}
\def\lmbh{log(M$_{\rm BH}$)}

\def\ledd{$\lambda_{\rm Edd}$}

\def\fobs{$F^{obs}_{14-195 {\rm keV}}$}
\def\lx{L$_{\rm X}$}
\def\lbol{L$_{\rm bol}$}
\def\snxv{$\sigma^2_{NXV}$}
\def\asnxv{$\overline{\sigma^2_{NXV}}$}

\begin{document}

   \title{The X-ray variability of AGN: power-spectrum and variance analysis of the \swift/BAT light curves}

   \author{I. E. Papadakis \inst{1,2}
          \and
          V. Binas-Valavanis\inst{1,2}
          }

    \titlerunning{X-ray AGN PSDs}
   \authorrunning{I. E. Papadakis \& V. Binas-Valavanis}

   \institute{Department of Physics and Institute of Theoretical and Computational Physics, University of Crete, 71003 Heraklion, Greece \email{jhep@physics.uoc.gr}
         \and
             Institute of Astrophysics, FORTH, GR-71110 Heraklion, Greece\\
             }

   \date{}

 
  \abstract
  {}
  {We study the X--ray power spectrum of Active Galactic Nuclei (AGN) to investigate whether Seyfert I and II power spectra are similar or not, whether the AGN variability depends on black hole mass and accretion rate, and to compare the AGN power spectra with the Galactic X--ray black hole binaries power-spectra.}
   {We used 14-195 keV band light curves from the 157th \swift/BAT hard X--ray survey and we computed the  mean power spectrum 
   and excess variance of AGN in narrow black-hole mass/luminosity bins. We fitted a power-law model to the AGN power spectra, and we investigated whether the power spectrum parameters and the excess variance depend on black hole mass, luminosity and accretion rate.}
   {The Seyfert I and Seyfert II power spectra are identical, in agreement with AGN unification models. The mean AGN X--ray power spectrum has the same, power-law like shape with a slope of $-1$ in all AGN, irrespective of their luminosity and BH mass. We do not detect any flattening to a slope of zero at frequencies as low as $10^{-9}$ Hz. We detect an anti-correlation between the  PSD amplitude and the accretion rate, similar to what has been seen in the past in  the 2--10 keV band. This implies that the variability amplitude in AGN decreases with increasing accretion rate. The universal AGN power-spectrum is consistent with the mean, 2--9 keV band Cyg X-1 power spectrum in its soft state. We detect a small difference in amplitude, but this is probably due to the difference in energy.}
   {The mean, low frequency AGN X--ray power spectrum is consistent with the extension of the mean, 0.01--25 Hz Cyg X-1 power spectrum in its soft state to lower frequencies. We cannot prove that the mean AGN PSD is analogous of the mean Cyg X-1 PSD in its soft state, as we do not know the location of the high frequency break in the hard X-ray AGN PSDs. However, if that is the case, then the accretion disc in AGN probably extends to the radius of the innermost circular stable orbit (as is probably the case with the black hole binaries in their soft sate). The X-ray corona will then be located on top, illuminating the disc and producing the X--ray reflection and disc reverberation phenomena that are commonly observed in these objects. Furthermore, the agreement between the AGN and the Cyg X--1 power spectrum (either in the soft or the hard state) over many decades in frequency indicates that the X--ray variability process is probably the same in all accreting objects, irrespective of the mass of the compact object. We plan to investigate this issue further in the near future.}


   \maketitle
%

\section{Introduction}

Active Galactic Nuclei (AGN) vary at all wavebands, with the variability amplitude increasing with increasing energy. X--rays vary significantly in flux and/or spectral shape on time scales as short as minutes/hours in some cases. This supports the hypothesis that X--rays are emitted in a small region located close to the super-massive black hole (BH), which resides in the center of these objects. X--ray spectral and timing studies are though to be important as their study could help us understand the physical processes that operate in the innermost region of these objects. 

AGN variations are stochastic in nature, therefore their study involves the use of statistical tools like the so-called normalized excess variance, \snxv, and the power-spectral density function (PSD). The former is equal to the integral of the latter, hence it provides limited information on the nature of the variability when compared to the study of the PSD itself. Power spectrum analysis of the X-ray light curves of radio-quiet AGN was first introduced with the use of long, EXOSAT light curve thirty years ago \citep[e.g.][]{Lawrence93,McHardy93}. Detailed characterization of the X--ray power spectra of AGN (in the 2--10 keV band) was later achieved with the combination of RXTE and {\it XMM-Newton} light curves \citep[e.g.][]{Uttley02,Papadakis02,Markowitz03,McHardy04,Uttley05,Markowitz07,Markowitz10,Gonzalez-Martin2012}. The past variability studies have established that the X--ray PSDs in the 2--10 keV band have a featureless, power-law like shape with a slope of $\sim -1$ over many decades in frequency, which then steepens to a slope of $\sim -2$ (or even steeper) above a bend-frequency, $\nu_b$. This characteristic frequency depends on BH mass and possibly on the accretion rate rate as well \citep{McHardy06,Gonzalez-Martin2012}. This power spectrum shape is similar to the shape of the X--ray power spectra of the Galactic X--ray BH binaries (GBHs) in their soft states.

We use the \swift/BAT light curves from the 157 month BAT survey (Lien et al, in preparation) of a large sample of AGN, in order to estimate their power spectrum at low frequencies. BAT has been continuously monitoring the sky for $\sim 18.5$ years, and the 157 month survey provides long, evenly sampled, and high signal-to-noise light curves for a large number of AGN in the 14-195 keV band. This energy band is well suited for the study of the intrinsic X--ray variability in AGN, because it is unaffected by any potential absorption variations (as long as the intrinsic hydrogen column density is less than a few $\times 10^{24}$ cm$^{-2}$). 

\swift/BAT light curves have been used in the past to study the flux and spectral variability of AGN \citep[e.g.][]{Beckmann07,Caballero12,Soldi14}. Our work is similar to the work of \cite{Shimizu13}. These authors (SM13, hereafter) calculated the power spectrum of 30 AGN, for the first time in hard X-rays (i.e. at energies higher than 10 keV). They used 58 month long BAT light curves, and they measured the PSD at frequencies $\sim 10^{-8}-10^{-6}$ Hz. They found that all power-spectra were well fitted by a single power-law model with a slope of $\sim -1$, and they did not find any significant correlation between the PSD parameters and various AGN properties, including luminosity and BH mass (\mbh).

Our sample consists of the one hundred, X-ray brightest, radio quiet AGN from the BASS DR2 survey\footnote{https://www.bass-survey.com/} \citep{Koss2022}. Our main aim is to use the high quality, \swift/BAT light curves to compute their long term power-spectrum and:  1) determine whether there are any differences between the average PSD of Seyfert 1 and Seyfert 2s, 2) investigate if and how does the PSD vary with AGN properties (like BH mass and luminosity), and 3) compare the mean AGN PSD with the PSD of GBHs. 

\section{The sample}
\label{sec:sample}

\cite{Koss2022} presented a catalog of AGN for the second data release of the Swift BAT AGN Spectroscopic Survey (BASS). They have identified all AGN among the 1210 sources in the BAT 70-month survey \cite{Baumgartner2013} and have obtained BH mass estimates for almost all of them. Our sample consists of the 100 brightest AGN (in the 14-195 keV band) in Table 6 of \cite{Koss2022}, with the exception of Q0241+622, because it lacks a black hole mass estimate. In choosing the sources we excluded: a) "beamed" AGN (i.e. objects classified as "BZQ", "BZG" and "BZU" by \cite{Koss2022}) since their emission may be dominated by the jet emission, and b) "dual" AGN (listed in Table 4 of \cite{Koss2022}), if the ratio of the predicted \fobs\ of the fainter source over the total detected BAT flux is larger than 1\% (see \S2.3 in \cite{Koss2022}). 

The sources in our sample are listed in Table \ref{table:sampledata}. Source classification, redshift, and logarithm of black hole mass, \lmbh, are listed in columns 2, 4, and 5, and are taken from \cite{Koss2022}. In the third column we list the mean observed flux in the 14-195 keV, \fobs, taken from the \swift/BAT, 157-month, hard X--ray survey web page\footnote{https://swift.gsfc.nasa.gov/results/bs157mon}.

In the sixth column we list the logarithm of the intrinsic X--ray luminosity, L$_{\rm X}$, in the 14--150 keV band. It is computed by dividing the bolometric luminosity (\lbol) of \cite{Koss2022} by 8 (i.e. the bolometric correction factor used by these authors). The 14-195 keV band X--ray luminosity should be a measure of \lbol, which is one of the important physical properties of an AGN (together with \mbh). Once the BH mass is known, the ratio \ledd=\lbol/L$_{\rm Edd}$ (where L$_{\rm Edd}$ is the Eddington luminosity), should be a rough estimate of the ratio of the accretion rate over the Eddington accretion rate. One of our objectives is to examine the dependence of the PSD on \ledd. However, the bolometric conversion factor between \lx\ and \lbol\ is uncertain. For example, \cite{Koss2022} assume \lbol/L$_{\rm 14-195 keV}$=8, while \cite{Beckmann09} assumed \lbol/L$_{\rm 20-100 keV}$=6. In addition, the conversion between \lx\ and \lbol\ may be more complicated than just applying a constant bolometric correction \citep[e.g.][]{Lusso12}. For this reason, we list the  \lx\ measurements in Table \ref{table:sampledata}, and we will consider the dependence of the PSD on both \lx\ and \ledd\ in our study below. The AGN in our sample span a range of $\sim 3$ in \lmbh\ (from $\sim 10^6$ to $\sim 10^9$ M$_{\odot}$) and a range of $\sim 2.5$ in log(\lx) (from $\sim 10^{42}$ to $\sim 10^{44.5}$ ergs/s). They should be representative of the radio-quiet AGN population in the nearby Universe.

\section{The light curves}

We retrieved BAT light curves for all the sources in the sample from the Swift-BAT 157 month survey web page. We considered the monthly, Crab-weighted light curves in the 14-195 keV band (Lien et al, in preparation). The 157 month light curves are constructed as in the 105 month release \citep{Oh2018}, with some updates in both the instrumental calibration and responses (A. Lien, priv. comm.). 

A few points in some light curves are associated with large error bars. 
Their exposure time is less than 5 ksec (a small number when compared to the nominal bin width of 1 month). We use the mean of the error squared to compute the Poisson noise level in the power spectrum (see \S\ref{sec:noiseestimation}). Since the error of these points can bias the mean squared error, we decided to remove these points and to replace them using linear interpolation between the adjacent points in the light curves. 
There are also a few missing points in some light curves, which we also fill in using linear interpolation. We added a random error to each interpolated point, assuming Gaussian statistics with a standard deviation equal to the mean error of all points in the light curves (excluding the ones with $\Delta t \le 5$ ks). There are 39 light curves with missing points and/or points with an exposure time less than 5 ks (the mean number of missing points is 2).

We checked which sources show significant variations in the light curves using traditional $\chi^2$ statistics. To this end, we computed the weighted mean of each light curve\footnote{The count rate error in the \swift-BAT light curves is inversely proportional to the square root of the exposure time in each bin (this is easily checked when plotting errors versus the respective exposure time), and it does not depend on the source's flux. In this case, the weighted rather than the straight mean is more representative of the source's mean flux.}, and we fitted the light curves with a constant line, equal to the mean. The fit was done without considering the interpolated points. The resulting $\chi^2$ values over the degrees of freedom (dof) are listed in the 7th column in Table \ref{table:sampledata}. We accept that a light curve is variable if the null hypothesis probability, $p_{null}$, is less than 0.01 (the null hypothesis is that the source is not variable). Letters "V" or "NV" in the 8th column in Table \ref{table:sampledata} indicate whether a source is variable or not (i.e. $p_{null}<0.01$ or $p_{null} \ge 0.01$, respectively). 

Twenty six sources out of the 100 in the sample are non-variable. Nineteen of the NV sources are Sy1 and the rest are Sy2 and Sy 1.9. The mean \lmbh\, and log(\lx) are $\sim 7.8$ and $\sim 43.6$, respectively, for both the V and NV group of sources. However, the ratio of the mean count rate over the mean error (i.e. the signal-to-noise ratio of the light curve, S/N) is smaller than  3 in almost all (but two) of the NV sources. On the contrary, the same ratio is larger than 3 in almost two thirds of the V sources. This result indicates that the reason we do not detect significant variations in the NV sources is mainly because their S/N ratio is low. As a result, the amplitude of the variations due to the experimental noise is larger than the amplitude of the intrinsic variations, and we cannot detect them.

\section{Power spectrum estimation}
\label{sec:psdestimation}

As it is often the case, we used the periodogram as an estimate of the intrinsic power spectrum. We computed the periodogram of each light curve in the usual way, i.e.

\begin{equation}
I_{N}(f_j)  =  \frac{2\Delta t_{rf}}{N} |\sum^{N}_{i=1} x(t_i)e^{-i2\pi f_{j,rf}t_i}|^2,
\label{periodogram}
\end{equation}

\noindent 
where $\Delta t_{rf}$ is the rest frame light curve bin size ($\Delta t_{rf}=1{\rm month}/(1+z); z$ is the source redshift), $N$ is the total number of points in the light curve, and the data are normalized to the mean (i.e. $x(t_i)=[x(t_i)-\overline{x}]/\overline{x}$, with $\overline{x}$ being the light curve mean). The periodogram is computed at the usual set of frequencies, i.e. $f_{j,rf}=j/(N\Delta t_{rf}), j=1,2,...,N/2$ ($N=157$ and $j_{max}=78$).

\subsection{Computing the Poisson noise error}
\label{sec:noiseestimation}

In order to estimate the intrinsic PSD, we subtracted the constant Poisson noise level, $C_{PN}$, from the periodogram. We computed $C_{PN}$ as follows, 

\begin{equation}
C_{PN}  =  \frac{2\Delta t_{rf} (\overline{\sigma^2_{err})}}{\overline{x}^{2}},
\label{eq:psdnoiselevel}
\end{equation}

\noindent where $\overline{\sigma^2_{err}}$ is the mean of the squared error of the points in the light curve. This should be a good estimate of the light curve variance due to the Poisson noise. Since the correct estimation of $C_{PN}$ is important for the accurate determination of the PSD, we used the power spectrum of the NV sources to test eq.\, (\ref{eq:psdnoiselevel}).

The periodograms of the NV sources appear to be flat, as expected. To verify this, we computed the logarithm of the periodogram, and we binned them into groups of size $M=20$. The resulting PSD estimates are approximately Gaussian distributed, with known error \cite[][PL93 hereafter]{Papadakis93}. A constant line fits well all the binned logarithmic periodograms. This confirms that the power spectra of the NV sources are flat, indicating the lack of intrinsic variations with amplitude larger than the Poisson noise variations in the NV sources. In this case, the mean of the logarithm of the periodogram estimates (plus 0.25068; see \cite{Vaughan05}) should be representative of the logarithm of the actual Poisson noise level in the light curves, ${\rm log}(C_{PN,obs})$\footnote{We computed the mean of log[$I_N(f_j)$] instead of the mean of $I_N(f_j)$  
because the error of the latter is known, and is equal to $\sqrt{0.31/j_{\max}}$ (PL93).}.

\begin{figure}
   \centering
   \includegraphics[width=10cm]{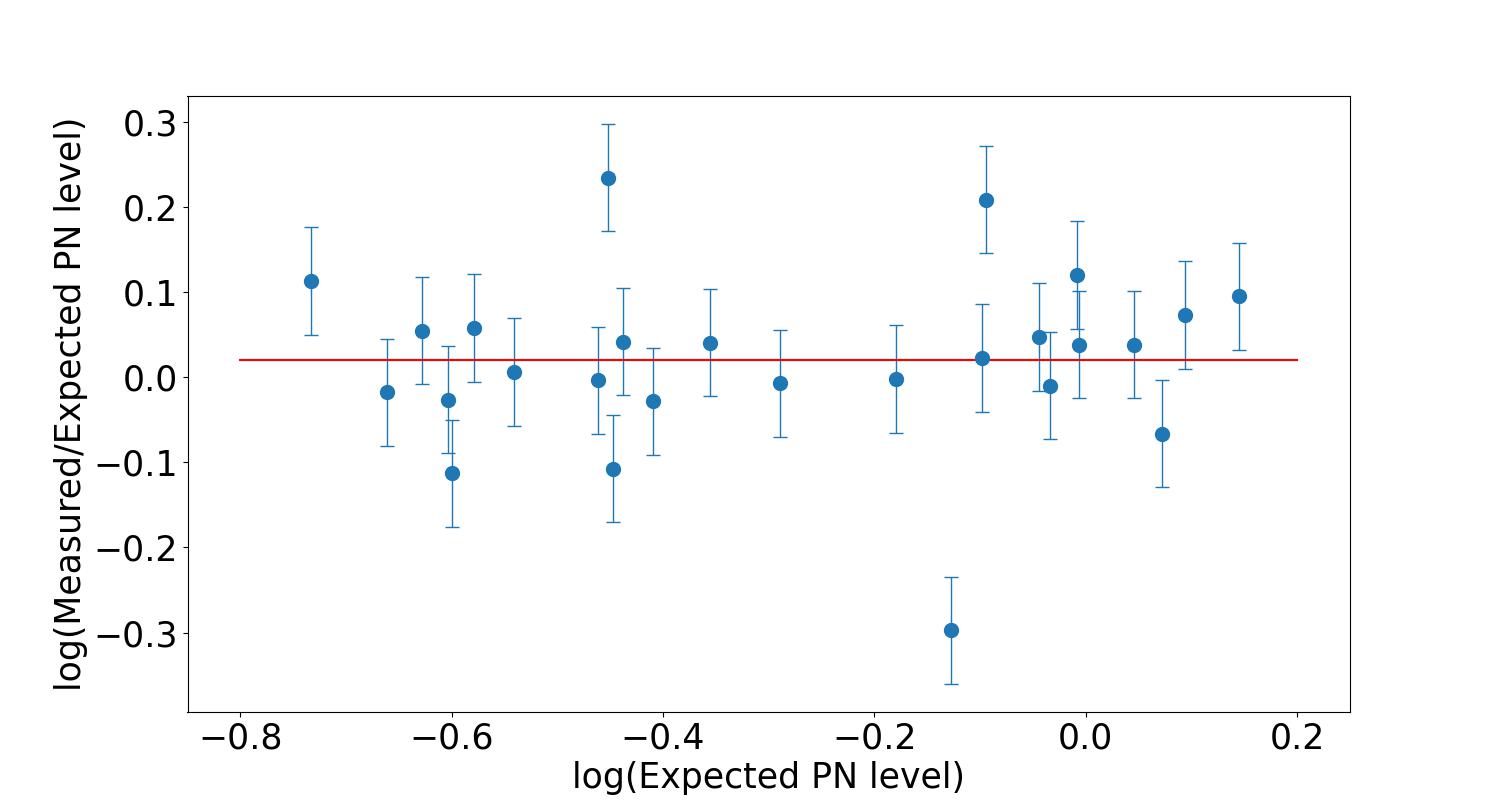}
      \caption{Plot of log$(C_{PN,obs}/C_{PN})$ versus $C_{PN}$ for the 26 NV sources in the sample. The solid line indicates the mean log$(C_{PN,obs}/C_{PN})$.}
         \label{fig:pnoiseratio}
\end{figure}

Figure \ref{fig:pnoiseratio} shows a plot of the logarithm of the ratio of $C_{PN,obs}$ over $C_{PN}$ (computed using eq.\,\ref{eq:psdnoiselevel}) versus log$(C_{PN})$, for the NV sources. The solid line shows the mean log($C_{PN,obs}/C_{PN})$, which is equal to $0.020\pm0.012$. The data are broadly consistent with this line, and since the mean log($C_{PN,obs}/C_{PN})$ is also consistent with zero (within $\sim 1.6\sigma$), we conclude that $C_{PN}$, as defined by eq.\,(\ref{eq:psdnoiselevel}), provides a good estimate of the Poisson noise in the light curves. 

Although the solid line in Fig.\,\ref{fig:pnoiseratio} appears to agree with data rather well, the best-fit $\chi^2$ value of 63.2 for 25 dof indicates that the statistical quality of the fit is rather poor. However, this is due to two sources mainly, where log($C_{PN,obs}/C_{PN})$ is $\sim 3-4\sigma$ away from the mean. If we do not consider these objects, the agreement between $C_{PN,obs}$ and $C_{PN}$ is quite good ($\chi^2=30.3/23$dof, $p_{null}=0.14$). We conclude that, although $C_{PN}$ may not (always) provide an accurate measurement of the Poisson noise on an individual light curve basis, it gives a good estimate of $C_{PN,obs}$ on average. Since our work is based on the study of the average PSD of many sources in various groups (see next Section), we adopt $C_{PN}$, as defined by eq.\,(\ref{eq:psdnoiselevel}), as an estimate of the Poisson noise in the \swift/BAT light curves.

\subsection{Computing and fitting the ensemble PSD of AGN}
\label{sec:psdfit}

Although the periodogram is an unbiased estimator of the intrinsic PSD, its statistical properties are not ideal. The probability distribution of the periodogram estimates follows a $\chi^2$ distribution with 2 degrees of freedom, their variance is large and unknown. In fact, the variance does not even decrease with increasing data points. For that reason, it is customary to smooth the periodogram using various ``spectral windows". However, smoothing in the linear space is not ideal in the case of power-law like intrinsic PSDs (PL93). One option would be to bin the logarithmic periodogram as suggested by PL93, but that is not the best solution in our case either. Since $j_{max}=78$, we would end up with just four points in the power spectrum, over a limited frequency range, if we would follow the prescription of binning the log periodogram into groups of size $M=20$. Fitting them with a straight line would result in best-fit parameters with large uncertainties. To avoid this, we followed a different approach to compute and fit the power spectra of the variable sources in the sample.

As we describe in the next section, we consider AGN in relatively small boxes in the log(\lx) vs \lmbh\, plane (so that they have similar BH mass and luminosity). Suppose there are $n_{\rm AGN}$ in such a box (typically, $n_{\rm AGN}=4-6$).
We compute the periodogram of each one of them and we subtract the Poisson noise level. Then we compute the mean of the  $n_{\rm AGN}$ periodograms, say $\overline{I(\nu)}$, at frequencies lower than $0.1$ month$^{-1}$, and we accept log$[\overline{I(\nu)}]$ 
as an estimate of the ensemble, mean power spectrum of the AGN in each [log(\lx),\lmbh] box. In a few cases, $\overline{I(\nu)}$ is negative at one or more frequencies. In these cases, we group these values together with the two $\overline{I(\nu)}$ in the adjacent frequencies, and we replace all three  with their mean. 

At higher frequencies, the variance of the noise subtracted power spectrum increases. In order to reduce the noise of the PSD at these frequencies, we further smooth the mean $\overline{I(\nu)}$'s  using a simple top-hat window of size 5, i.e. we group every 5 successive $\overline{I(\nu)}$'s together, we compute their mean and accept its logarithm as an estimate of the intrinsic PSD at the mean of the respective frequencies. 




Since 
$n_{\rm AGN}$ is small, the probability distribution of the resulting PSD estimates
is not Gaussian, even at high frequencies. Therefore we cannot use traditional $\chi^2$ statistics to fit the observed power spectra. For that reason, we assumed that the intrinsic PSD follows a power-law like shape, and  we fitted the power spectra, in the log-log space, with a line of the form, 


\begin{equation}
{\rm log(PSD)}(\nu)=\log(PSD_{amp})+PSD_{slope}\cdot {\rm log}(\frac{\nu}{\nu_0}),
\label{eq:linemodel}
\end{equation}

\noindent where $PSD_{amp}$ is the PSD amplitude at $\nu_0=10^{-2}$ month$^{-1}$. We chose this frequency to define the PSD amplitude because it is within the range of sampled frequencies and it is reasonably low, hence the best-fit value will not be affected by aliasing (or other) effects (see the discussion below).
We fitted the data following the ordinary least square (OLS) [Y|X] prescription of \cite{Isobe90}. In this way, even though we do not know the error of the logarithmic estimates of the power spectrum, we can still compute the best-fit line parameters, together with their error. 

\begin{figure}
   \centering
   \includegraphics[width=10cm]{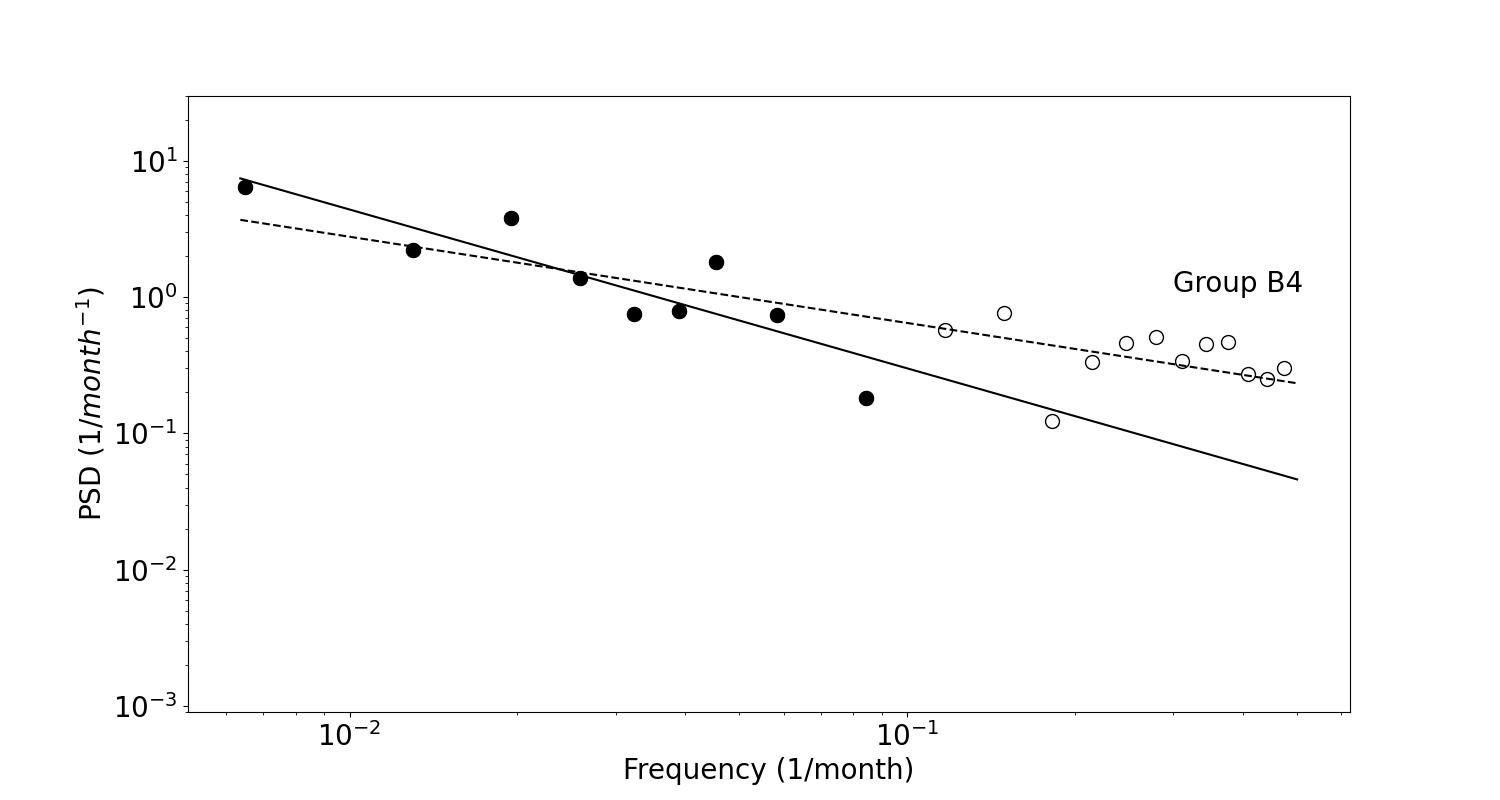}
      \caption{The mean power spectrum of the AGN which belong to group B4, as indicated in Fig.\,\ref{fig:lxvsbhmvar}. The power spectrum is computed as explained in \S\ref{sec:psdestimation}. Solid and dashed lines show the best-fit lines when we fit the low frequency and the full range power-spectrum (filled and filled+open circles, respectively), with the model defined by eq.\,\ref{eq:linemodel}. }
         \label{fig:examplepsd}
\end{figure}

As an example, Fig.\, \ref{fig:examplepsd} shows the average PSD of 4 AGN with  \lmbh\ $\sim7$ and log(\lx) $\sim 43.5$ (these are the objects that belong to the AGN group B4 in Fig.\,\ref{fig:lxvsbhmvar}). The solid and dashed lines show the best-fit lines when we fit eq.\,\ref{eq:linemodel} to the PSD at frequencies below 0.1 months$^{-1}$ and at all frequencies, respectively. The full-range best-fit slope is flatter than the low-frequency slope, with $\Delta (PSD_{slope})\sim 0.5$. This is the PSD with one of the largest differences between the low and the full frequency range best-fit slopes. Differences between the low-frequency and the full range best-fit slopes appear in other power spectra as well, but they are smaller than 0.5 (see Table \ref{table:psdsloperes}).

A PSD flattening may be due to the smoothing process at high frequencies, which may bias the observed power spectrum and it may appear flatter than the intrinsic PSD. In addition, \swift/BAT does not observe continuously all AGN. In quite a few cases, the total exposure time in some bins is less than a day, which is much smaller than the bin size of the light curves. It is possible then that aliasing effects may be significant, in which case they will also flatten the observed PSDs at high frequencies. Such effects are difficult to model (as they depend on the intrinsic PSD and the actual observing pattern within each $\Delta t$). For that reason, we fitted the PSDs twice: once at frequencies below $10^{-1}$ month$^{-1}$ (the LF fits, hereafter), and the second time over the full frequency range (the FR fits, hereafter).

\begin{figure}
   \centering
   \includegraphics[width=10cm]{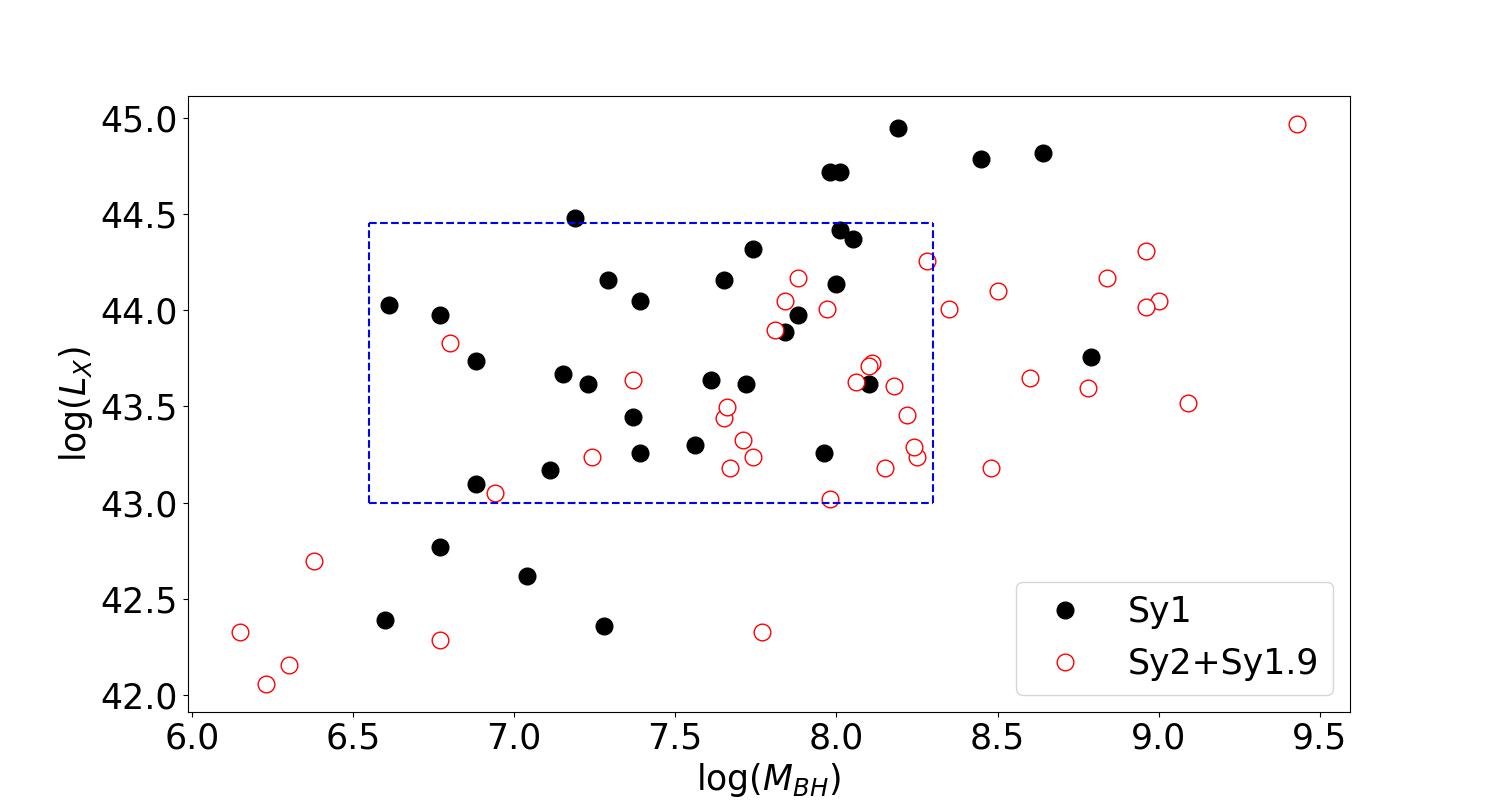}
   \includegraphics[width=10cm]{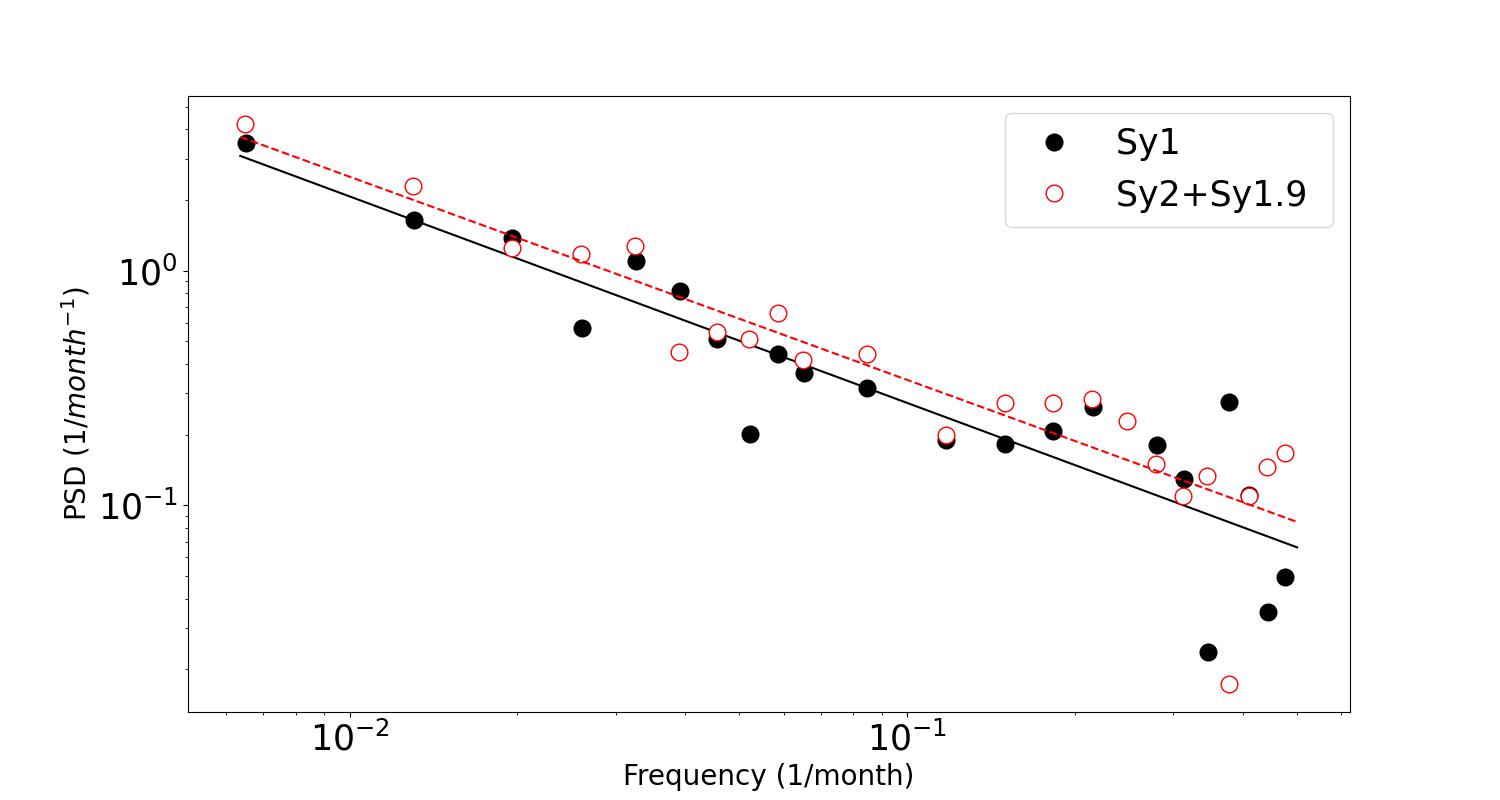}
      \caption{The upper panel shows a plot of log(\lx), versus \lmbh\, for the variable sources in the sample. Black and red circles indicate data for the Sy1 and Sy2 galaxies, respectively. The mean BH mass and \lx\ of the Sy1 and Sy2 sources in the dashed box are similar. {\it Lower panel:} Black and red points show the ensemble PSD of the Sy1 and Sy2 galaxies which are in the dashed box drawn in the upper panel. Black and red dashed lines show the best-fit line to the PSDs. }
         \label{fig:sy1vssy2}
\end{figure}

\section{Power spectrum analysis}

\subsection{The Sy1 and Sy2 PSDs}

We first investigated whether the Sy1 and Sy2 power spectra are the same or not. The top panel in Fig.\,\ref{fig:sy1vssy2} shows a plot of log(\lx) versus \lmbh\, for all the variable sources in the sample. Black filled and red open circles indicate the Sy1 and the Sy2 sources, respectively (hereafter, we refer as ``Sy2" both the Sy1.9 and Sy2 sources in the sample). On average, \mbh\ in Sy2s appears to be larger when compared to Sy1s, while their intrinsic X--ray luminosity appears to be smaller. The investigation of the significance of these differences is outside the scope of our work, however, since the PSD may depend on \mbh\, and/or \lx\, we need to compare the mean power-spectrum of Sy1 and Sy2 with the same \mbh\ and \lx\ in order to test if they are similar or not. 

There are 23 Sy1 and 23 Sy2s in the box indicated by the dashed lines in the upper panel of Fig.\,\ref{fig:sy1vssy2}. Their mean \lmbh\ and log(\lx) are consistent (within the errors): $\overline{\rm log(M_{BH}})=7.5 \pm0.1$, and $7.8\pm 0.1$ for Sy1 and Sy2, while $\overline{\rm log(L_{X})}=43.8\pm 0.1$ and 43.6$\pm 0.1$, respectively. 

The lower panel in Fig.\,\ref{fig:sy1vssy2} shows the mean PSD of all the Sy1 and Sy2 AGNs within the same box (black filled, and red open circles, respectively). We fitted the power spectra with the model defined by eq.\,(\ref{eq:linemodel}). Black solid and red dashed lines show the best-fit lines to the full range Sy1 and Sy2 PSDs, respectively. 
The LF(FR) best fit line parameters are: log$(PSD_{amp,\rm Sy1})=0.36\pm 0.26(0.32\pm 0.28$), log($PSD_{amp,\rm Sy2})=0.43\pm 0.31(0.40\pm0.32$), and $PSD_{slope,\rm Sy1}=-1.00\pm 0.07(-0.88\pm 0.09), PSD_{slope,\rm Sy2}=-0.95\pm 0.05(-0.87\pm 0.09$). Clearly, the Sy1 and Sy2 PSDs are consistent, within the errors. This result suggests that the variability properties of the Sy1 and Sy2 galaxies in the 14-195 keV band are identical. Given this result, we combine together light curves from Sy1 and Sy2 sources in order to investigate  the dependence of the power spectrum on BH mass and/or luminosity. 

\begin{figure}[h]
  \centering
  \includegraphics[width=10cm]{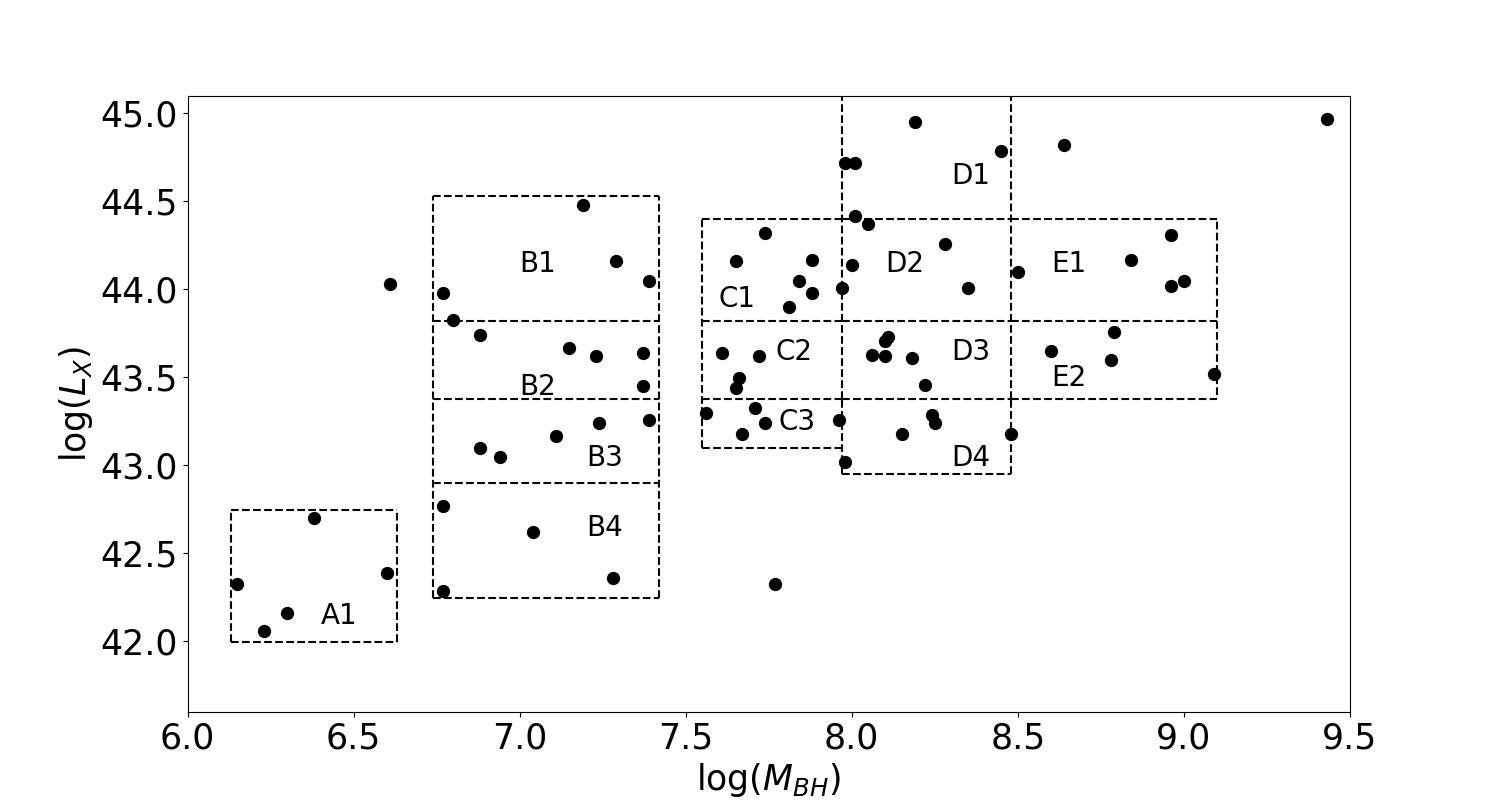}
     \caption{Plot of log(\lx), versus \lmbh\, for the variable sources in the sample. AGN within the boxes indicated by the dashed lines have similar \mbh\ and \lx.} 
        \label{fig:lxvsbhmvar}
\end{figure}

\subsection{The dependence of PSD slope on \mbh, \lx, and \ledd.}
\label{sec:psdsloperes}

It is not easy to model the periodogram of each individual AGN in our sample and investigate if and how it may depend on \mbh\, and/or \lx. Instead, we grouped the variable AGN into relatively narrow BH mass and luminosity bins, as indicated by the dashed lines in Fig.\,\ref{fig:lxvsbhmvar}. Their size is small ($\Delta$log(\mbh)$\sim\Delta$log(\lx)$\sim 0.5$), so that \mbh\ and \lx\ of AGN in each box should be within a factor of $\sim 2$ or so. Consequently, their ensemble PSD should provide a good estimate of the intrinsic PSD of an AGN with BH mass and X--ray luminosity equal to the mean \mbh\ and \lx\ of the objects in each box.  

We computed the mean power spectrum of all AGN in each box (as explained in \S\ref{sec:psdestimation}), and we fitted them with a line, as defined by eq.\,\ref{eq:linemodel}. The best-fit slope should be indicative of the intrinsic slope of the AGN PSD. On the other hand, the best-fit amplitude may overestimate the intrinsic PSD amplitude, because we have considered only the variable AGN in Fig. \ref{fig:lxvsbhmvar}. This is appropriate when studying the PSD slope, as information on the slope can be provided only by sources with PSDs above the Poisson noise level. However, if there is a distribution of PSD amplitudes at each \mbh,\lx, then we should consider all AGN in each [\mbh,\lx] bin in order to measure the mean of the distribution, irrespective of whether they are variable or not (as long as they are at the same distance). We therefore use the best-fit slope from the model fits to the average PSDs of the variable AGN to study whether $PSD_{lope}$ depends on \mbh\, and/or \lx, and we will discuss the dependence of $PSD_{amp}$ on these parameters in the next Section. 

\begin{figure}
   \centering
   \includegraphics[width=9cm]{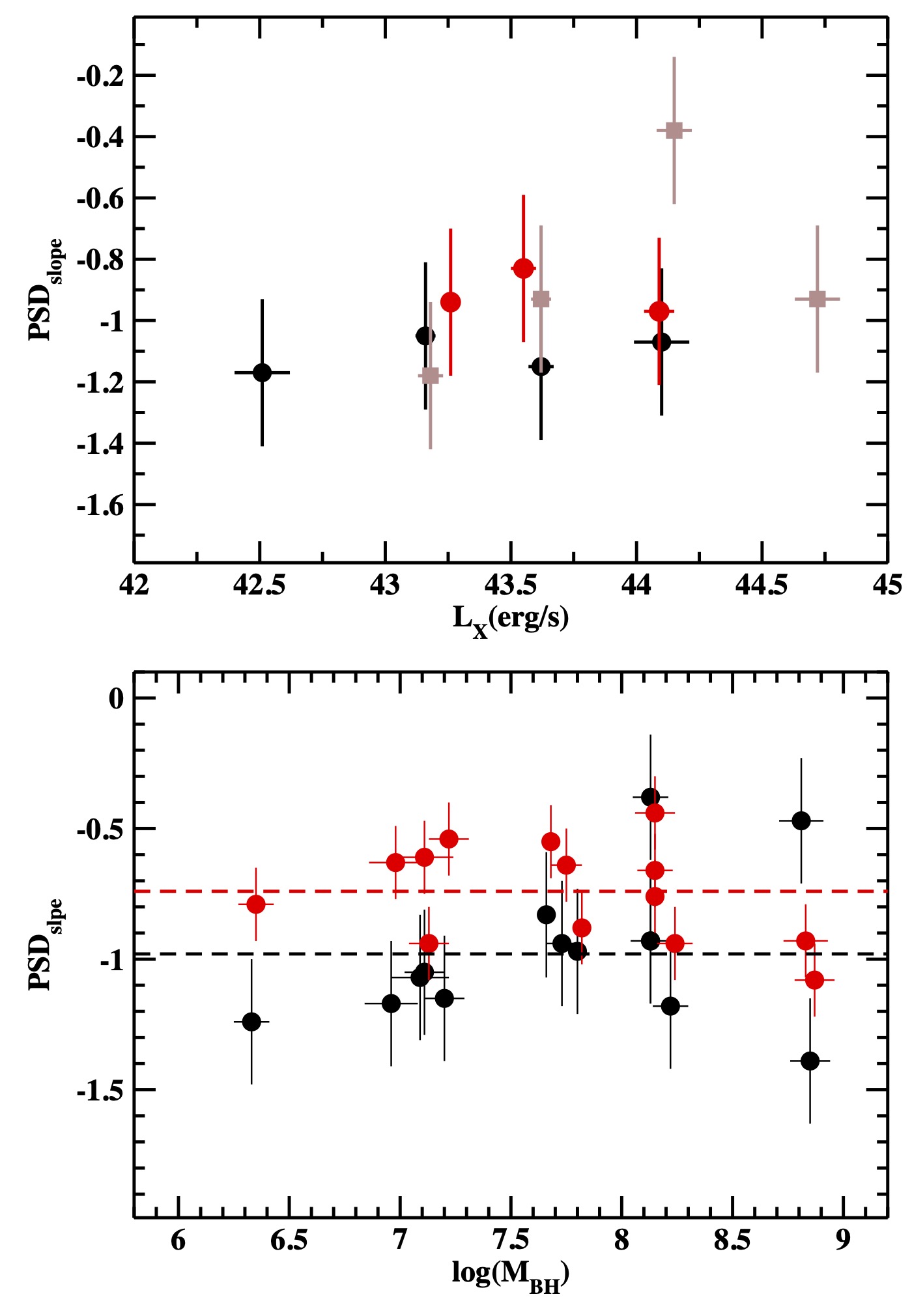}
      \caption{{\it Upper panel:} LF, best-fit $PSD_{slope}$ plotted as function of log(\lx) for the B, C and D groups (black circles, red circles, and brown squares, respectively). {\it Bottom panel:} The LF and FR $PSD_{slope}$ plotted as a function of $\overline{\log({\rm M_{BH}})}$ Black and red dashed lines indicate the mean LF and FR slopes, respectively.}
         \label{fig:res1}
\end{figure}

The mean \lmbh\, and log(\lx) of the AGN in each box in Fig.\,\ref{fig:lxvsbhmvar}, and their error, are listed in the first column of Table \ref{table:psdsloperes}. Second and third columns list the LF, and the FR best-fit PSD slopes, respectively. The top panel in Fig.\,\ref{fig:res1} shows the LF, best-fit PSD slope plotted as a function of the mean log(\lx) for the AGN in groups (B1, B2, B3 B4), (C1, C2, C3) and (D1, D2, D3, D4) (black circles, red circles and brown squares, respectively). All AGN in the B, C and D groups have similar BH masses, with a mean of $\sim 7.1, 7.7,$ and 8.1, respectively (see Table \ref{table:psdsloperes}). Within each of the B, C and D groups the PSD slope does not appear to depend on \lx. PSD slopes also appear to be similar between the various groups, which implies that $PSD_{slope}$ does not depend on \mbh\ either. 

\begin{table}[h]
\caption{First column lists the mean \lmbh, and log(\lx) of the objects in the groups labeled in Fig.\,~\ref{fig:lxvsbhmvar}. The LF and FR best-fit  $PSD_{slope}$ values are listed in the second and third columns, respectively. }
\begin{center}
\begin{tabular}{lcc}
\hline
\hline
\\
$ \overline{{\rm log(M_{BH})}},\overline{ {\rm log(L_X)} }$ & \multicolumn{2}{c}{$PSD_{slope}$} \\
 & (LF) & (FR) \\
$6.33\pm0.08,42.32\pm0.11$(A1) & $-1.24\pm0.36$ &  $-0.79\pm0.18$  \\
$7.09\pm0.13,44.10\pm0.11$(B1) & $-1.07\pm 0.32$ & $-0.61\pm 0.15$ \\
$7.20\pm0.09,43.62\pm0.05$(B2) & $-1.15\pm 0.47$ & $-0.54\pm 0.15$ \\
$7.11\pm0.09,43.16\pm0.04$(B3) & $-1.05\pm 0.20$ & $-0.94\pm 0.13$ \\
$6.96\pm0.12,42.51\pm0.11$(B4) & $-1.17\pm 0.20$ & $-0.63\pm 0.08$ \\
$7.80 \pm 0.04,44.09 \pm  0.06$(C1) & $ -0.97\pm 0.09$ & $-0.88\pm 0.11 $ \\
$7.66 \pm 0.02,43.55 \pm  0.05$(C2) & $-0.83\pm 0.23$ & $-0.55\pm 0.15$ \\
$7.73 \pm 0.07,43.26 \pm  0.03$(C3) & $-0.94\pm 0.18$ & $-0.64\pm 0.10$ \\
$8.13 \pm 0.09, 44.72 \pm 0.09$(D1) & $-0.93\pm 0.11$ & $-0.44\pm 0.09$ \\
$8.13 \pm 0.08, 44.15 \pm 0.07$(D2) & $ -0.38\pm 0.29$ & $-0.66\pm 0.26$ \\
$8.13 \pm 0.02,43.62 \pm  0.04$(D3) & $-0.93\pm 0.20$ & $-0.76\pm 0.10 $ \\
$8.22 \pm 0.08,43.18 \pm  0.05$(D4) & $-1.18\pm 0.24$ & $-0.94\pm 0.10 $ \\
$8.85 \pm 0.09,44.13 \pm  0.05$(E1) & $-1.39\pm 0.15$ & $-1.08\pm 0.09$ \\
$8.81 \pm 0.10,43.63 \pm  0.05$(E2) & $-0.47\pm 0.25$ & $-0.93\pm 0.21$ \\
\hline
\end{tabular}
\end{center}
\label{table:psdsloperes}
\end{table}

To investigate these issues quantitatively, we fitted the data plotted in the top panel in Fig.\,\ref{fig:res1} with a line of the form: log(PSD$_{slope})=a+b$[log(\lx)$-43.5$]. We took into account errors in both coordinates, using the {\tt fitexy} routine of \cite{Press92}. We expect the $PSD_{slope}$ error to be similar in all of them, since the mean PSD is computed in the same way and the number of objects is roughly the same in all bins. The errors listed in Table \ref{table:psdsloperes} fluctuate mainly because of the small number of points in the PSDs (specially in the low frequency part). The error on the slope turns out small in cases where the PSD points happened to align well around the best-fit line, while larger errors result in the case when the scatter of the points around the best-fit line is large (see also discussion in \S IV in \cite{Isobe90} on the best-fit errors in the case of small number of data points). For that reason, we computed the mean error of all PSD slope estimates,  and we assigned it to each $PSD_{slope}$, assuming it is more representative of their intrinsic error. 

The best-fit results from fitting the points in the top panel of Fig.\,\ref{fig:res1} are: $b_{B}=0.06\pm0.17, b_{C}=-0.08\pm0.19,$ and $b_{D}=0.08\pm0.14$. All three best-fit $b-$values are consistent with zero, which shows that $PSD_{slope}$ does not depend on \lx. Furthermore, $a_{B}=-1.08\pm0.16, a_{C}=-0.92\pm0.12,$ and $a_{D}=-0.98\pm0.14$. The best-fit $a$ values for the various groups are also consistent with each other, which implies that PSD$_{slope}$ does not depend on \mbh\ either.

The bottom panel in Fig. \ref{fig:res1} shows the best fit PSD slope as a function of \mbh\, for all groups shown in Fig.\,\ref{fig:lxvsbhmvar}, and listed in Table \ref{table:psdsloperes}. Black and red circles show the LF and FR PSD slopes, respectively. The mean LF and FR $PSD_{slope}$ is  $\overline{PSD_{slope,LF}}= -0.98 \pm 0.06$ and $\overline{PSD_{slope,FR}}=-0.74 \pm 0.04$. The black and red horizontal lines in the same panel indicate the mean LF and FR PSD slopes, respectively.  When fitting the LF and FR data with these lines we get $\chi^2=18.2/13$dof and 25.3/13dof, respectively. Both fits are statistically accepted ($p_{null}=0.15$ and 0.02, respectively).  

As a final test, we also investigated the possibility that the PSD slope may depend on accretion rate. We fitted the LF and the FR PSD slopes versus log(\ledd) with a straight line of the form: $PSD_{slope}[log(\lambda_{\rm Edd})]=a_{\rm \lambda_{Edd}}+b_{\rm \lambda_{Edd}}[{\rm log(\lambda_{Edd})+1]}$. The fits were done taking into account the error on both variables.  When considering the fit in the case of the LF slopes the improvement in the goodness of fit is not significant ($\Delta \chi^2=0.2$ for 1 degree of freedom). However, in the case of the FR slopes, a straight line improves the fit significantly ($\Delta \chi^2=9.3$ for one extra degree of freedom; $p_{null}=0.002$). This result implies that the PSD slope may depend on \ledd. The best-fit slope of the PSD slope versus $\lambda_{Edd}$ relation is $b_{\rm \lambda_{Edd,FR}}=0.19\pm0.06$, which implies that the PSD slope may flatten with increasing $\lambda_{Edd}$.

\subsection{The dependence of PSD amplitude on \mbh, \lx, and \ledd.}
\label{sec:psdampresults}

\begin{figure}[h]
  \centering
  \includegraphics[width=10cm]{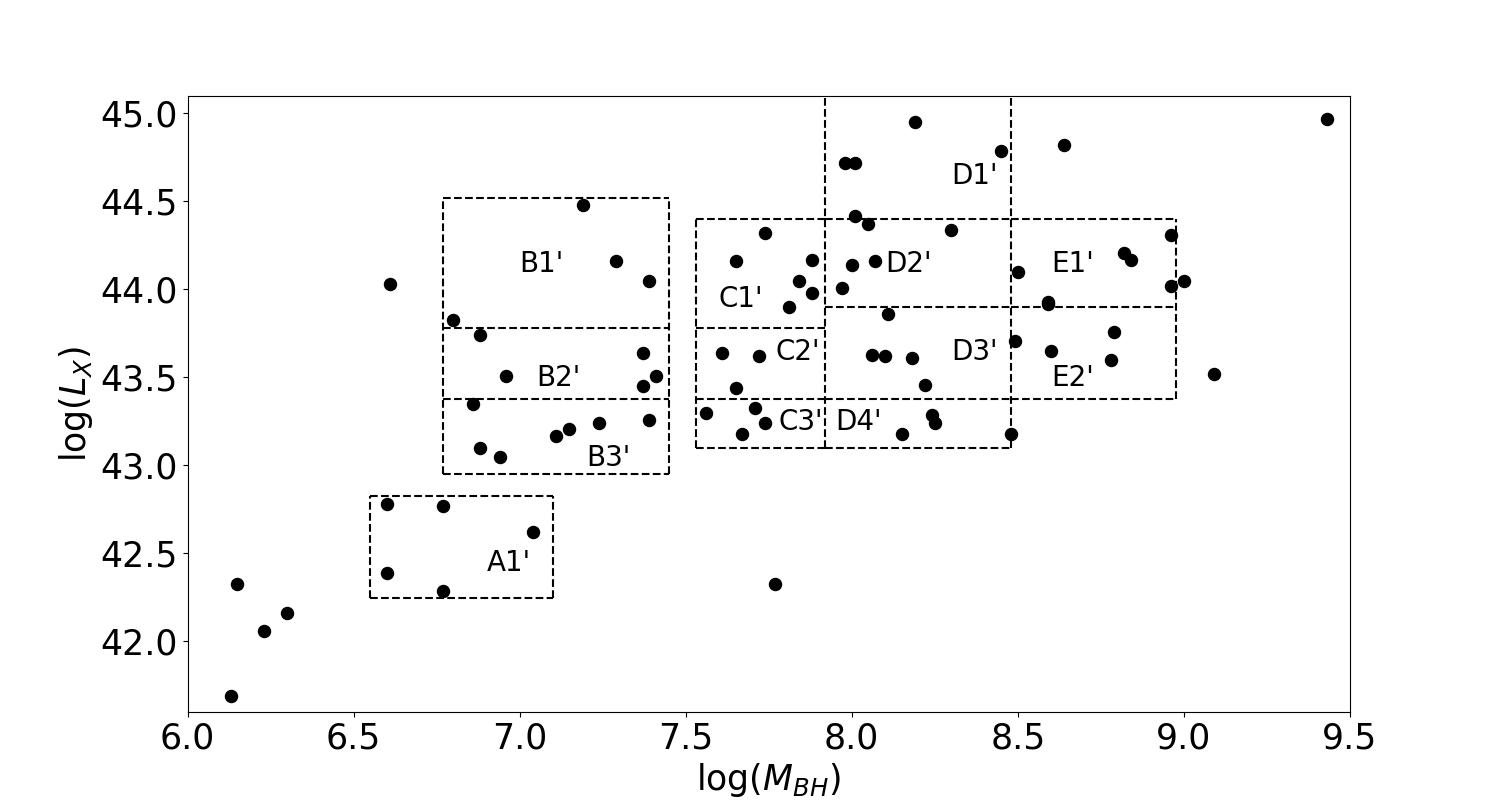}
     \caption{Like Fig.\,\ref{fig:lxvsbhmvar}, for AGN with S/N$\ge2$ light curves.} 
        \label{fig:lxvsbhmsnr2}
\end{figure}

As we discussed in the previous section, we should consider all AGN in each [\mbh,\lx] bin in order to measure the intrinsic PSD amplitude, irrespective of whether they are variable or not, as long as they are at the same distance. Distance determines the S/N ratio of the light curve for objects with the same luminosity. Even if the PSD amplitude is the same in two sources, if one of them is located far away the observed count rate will be small and its observed PSD may not be above the Poisson noise. The apparent difference in the observed PSDs in this case will be due to differences in the distance of the sources, and not in their PSD.   

For that reason, we considered all AGN in the sample with lightcurve with S/N$\ge2$. These objects are shown in Fig.\,\ref{fig:lxvsbhmsnr2}. Just like in Fig.\,\ref{fig:lxvsbhmvar}, the dashed lines in this figure indicate boxes with a small width so that all sources in each box should have similar \mbh\ and \lx. 
AS before, we computed the mean power spectrum of all AGN in each box in Fig.\,\ref{fig:lxvsbhmsnr2} as explained in \S\ref{sec:psdestimation}, and we fitted them with a line (as defined by eq.\,\ref{eq:linemodel}). The best-fit amplitude should be indicative of the (logarithm of the) intrinsic  PSD amplitude in AGN. The mean \lmbh\, and log(\lx) of the AGN in each box in Fig.\,\ref{fig:lxvsbhmsnr2}, together with their error, are listed in the first column of Table \ref{table:psdampres}. Second and third columns list the LF, and the FR best-fit PSD amplitudes, respectively. 

\begin{table}[h]
\caption{First column lists the mean \lmbh\ and log(\lx) of the objects in the boxes labeled in Fig.\,~\ref{fig:lxvsbhmsnr2}. The second and third columns list the LF and FR best-fit log($PSD_{amp})$ values, respectively.}
\begin{center}
\begin{tabular}{lcc}
\hline
\hline
\\
$ \overline{\rm log(M_{BH})},\overline{\rm log(L_X)}$ & 
\multicolumn{2}{c}{ $\log(PSD_{amp})$ } \\
 & (LF) & (FR) \\
$6.76 \pm 0.08$, $42.57 \pm 0.10$(A1$^{\prime}$) & $0.20\pm0.20$ & $-0.13\pm0.14$ \\
$7.17 \pm 0.13$, $44.13 \pm 0.14$(B1$^{\prime}$) & $0.00\pm0.11$ & $0.00\pm0.22$ \\
$7.20 \pm 0.11$, $43.57 \pm 0.05$(B2$^{\prime}$) & $-0.24\pm0.16$ &	 $-0.34\pm0.31$  \\
$7.08 \pm 0.07$, $43.19 \pm 0.04$(B3$^{\prime}$) & $0.17\pm0.13$ & $0.07 \pm0.13$ \\
$7.80 \pm 0.04$, $44.09 \pm 0.06$(C1$^{\prime}$) & $0.25\pm0.19$ & $0.21\pm0.21$ \\
$7.66 \pm 0.03$, $43.56 \pm 0.06$(C2$^{\prime}$) & $0.10\pm0.21$	& $-0.08\pm 0.15$ \\
$7.67 \pm 0.04$, $43.26 \pm 0.03$(C3$^{\prime}$) & $0.02\pm0.11$ & $-0.06\pm 0.14$ \\
$8.13 \pm 0.09$, $44.72 \pm 0.09$(D1$^{\prime}$) & $-0.02\pm0.12$ &	 $-0.22\pm0.17$ \\
$8.08 \pm 0.06$, $44.20 \pm 0.07$(D2$^{\prime}$) & $-0.23\pm0.46	$ & $-0.41\pm 0.38$ \\
$8.13 \pm 0.03$, $43.63 \pm 0.06$(D3$^{\prime}$) & $0.21\pm0.31$ & $0.06\pm0.12$  \\
$8.28 \pm 0.07$, $43.22 \pm 0.03$(D4$^{\prime}$) & $0.64\pm0.49$ & 		 $0.29\pm 0.28$ \\
$8.75 \pm 0.07$, $44.09 \pm 0.06$(E1$^{\prime}$) & $0.65\pm 0.40$ & $0.41\pm 0.43$ \\
$8.66 \pm 0.07$, $43.68 \pm 0.03$(E2$^{\prime}$) & $0.20\pm 0.14$ &	$0.34\pm 0.38$ \\
\hline
\end{tabular}
\end{center}
\label{table:psdampres}
\end{table}

\begin{figure}
    \centering
    \includegraphics[width=9cm]{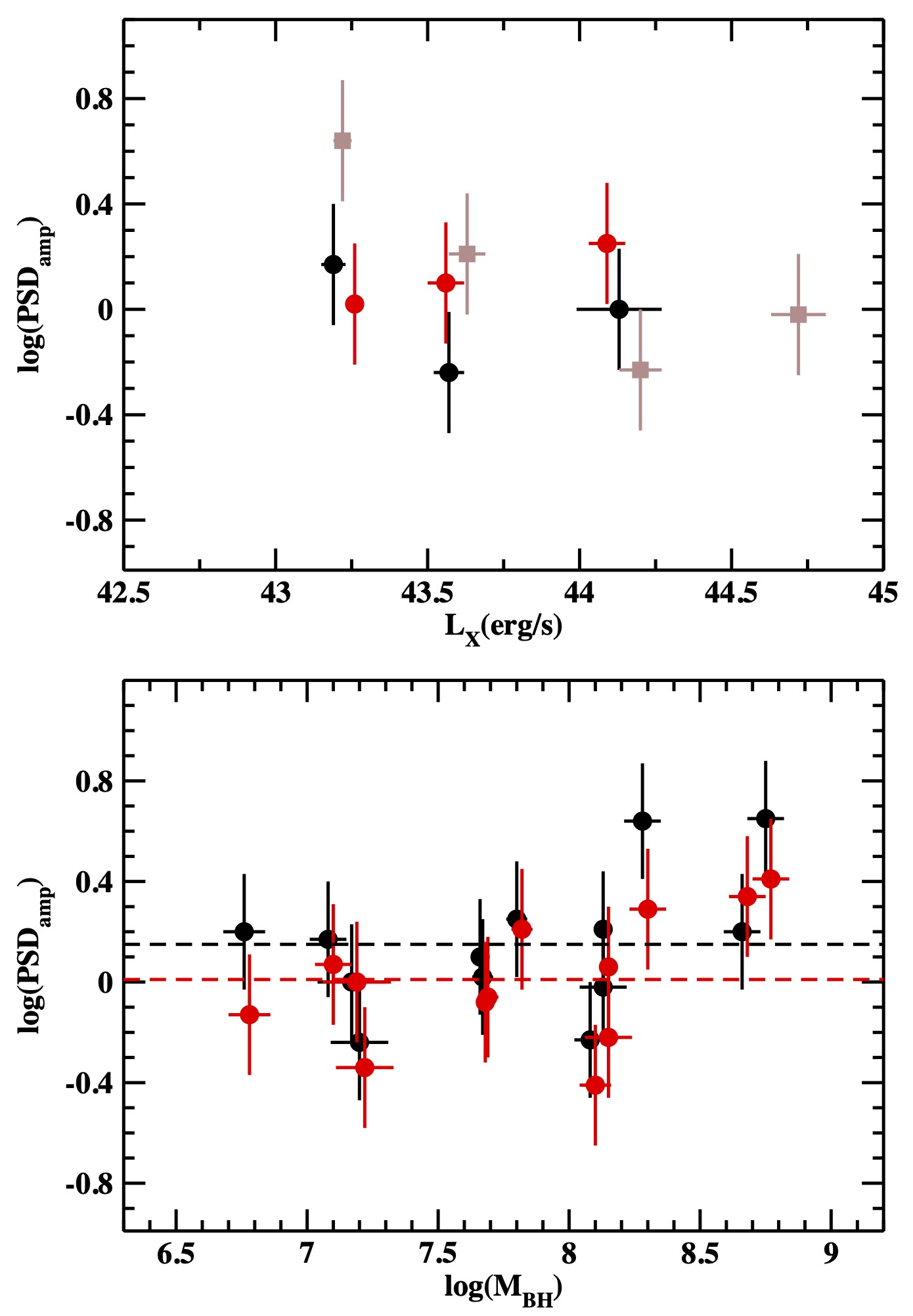}
     \caption{Same as in Fig.\,\ref{fig:res1}, for the PSD amplitude. The black, red, and brown points in the upper panel show the LF, best-fit log($PSD_{amp})$ values plotted as a function of \lx\, for AGN in the B$^{\prime}$, C$^{\prime}$ and D$^{\prime}$ boxes in Fig.\,\ref{fig:lxvsbhmsnr2}. The bottom panel shows all the LF and the FR PSD amplitudes (listed in Table \ref{table:psdampres}) plotted as a function of \mbh\, (black and red points, respectively; the dashed lines indicate the respective mean PSD amplitudes). } 
        \label{fig:res2}
\end{figure}

The top panel in Fig.\,\ref{fig:res2} shows the LF, best-fit log$(PSD_{amp}$) plotted as a function of the mean log(\lx) for the AGN in boxes (B1$^\prime$, B2$^\prime$, B3$^{\prime})$, (C1$^\prime$, C2$^\prime$, C3$^\prime$) and (D1$^\prime$, D2$^\prime$, D3$^\prime$, D4$^\prime$) (black circles, red circles and brown squares, respectively). We fitted the data with a line of the form: log($PSD_{amp})=c+d$[log(\lx)$-43.5$]. The best fit results are: $c_B=-0.005\pm0.08, c_C=0.09\pm0.11, c_D=0.36\pm0.22,$ and $d_B=-0.14\pm0.20, d_C=0.28\pm0.34, d_D=-0.47\pm0.32$. The best-fit $d$ values are all consistent with zero, indicating that the power-spectrum amplitude does not depend on \lx. The best-fit $c$ values appear to increase as we go from group B to group D, indicating an increase of $PSD_{amp}$ with increasing \mbh. In reality, since $c$ is equal to the PSD amplitude of an AGN with log(\lx)=43.5, such a dependence on \mbh\ would imply $PSD_{amp}$ increases with decreasing accretion rate. However, formally speaking, the best-fit {c} values are  consistent with each other(within 1.5$\sigma$), suggesting that this apparent dependence of $log(PSd_{amp})$ on \mbh\ is not significant.

The bottom panel in Fig.\, \ref{fig:res2} shows the best fit log($PSD_{amp})$ amplitude as a function of \mbh\ for all the groups shown in Fig.\,\ref{fig:lxvsbhmsnr2} and listed in Table \ref{table:psdampres}. Black and red circles show the LF and FR best-fit results, respectively. The black and red dashed lines indicate the LF and the FR mean logarithmic PSD amplitudes respectively ($\overline{\log(PSD_{amp,LF})}=0.15 \pm 0.06$ month$^{-1}$, and $\overline{\log(PSD_{amp,FR})}=0.01 \pm 0.06$ month$^{-1}$). These lines fit the data well ($\chi^2=16.1/12$dof, and $\chi^2=14/12$dof).

We also investigated the possibility that the PSD amplitude may depend on accretion rate. We therefore fitted the LF and the FR log($PSD_{amp}$) versus log(\ledd) data with a straight line of the form: log($PSD_{amp})[\log(\lambda_{\rm Edd})]=a_{\rm \lambda_{Edd}}+b_{\rm \lambda_{Edd}}[{\rm log(\lambda_{Edd})+1]}$. The best fit results are: $b_{LF,\rm \lambda_{Edd}}=-0.30\pm0.12$, and $b_{FR,\rm \lambda_{Edd}}=-0.26\pm0.12$.  The improvement in the goodness of fit when we consider the straight line with a non zero slope is significant in the case of the LF PSD amplitudes, and less so in the case of the FR amplitudes: $\Delta\chi^2=7.3$ and 6.1 for one extra degree of freedom, respectively ($p_{null}=0.007$ and 0.014). These results are indicative of an anti-correlation between PSD amplitude and accretion rate (with the variability amplitude decreasing with increasing \ledd). This is consistent with the dependence of the best-fit $c$ values on \mbh\ that we mentioned above. 

Our results so far show that the X--ray power spectrum in the 14-195 keV band is the same in all AGN: it has a power-law shape, with the same average slope and amplitude,  irrespective of their BH mass and luminosity. This result holds irrespective of whether we consider the LF or the FR best-fit results. The mean PSD slope when we consider the LR and the FR results is different at the 3.9$\sigma$ level: $\overline{PSD_{slope,LF}}-\overline{PSD_{slope,FR}}=-0.27 \pm 0.07$. There is also a difference between the mean LF and FR PSD amplitude, but it is small and not significant: $\overline{PSD_{amp,LF}}-\overline{PSD_{amp,FR}}=0.14 \pm 0.09$. We also find evidence that the PSD amplitude decreases with increasing \ledd. 


\subsection{Normalised excess variance analysis.}
\label{sec:snxvresults}

We performed an excess variance analysis of the \swift/BAT light curves. The variance of a light curve is equal to the integral of the power spectrum from a low to a high frequency \citep[e.g.][]{Allevato13}. It  can be measured relatively easy, and can be used to verify the power spectrum results that we report in the previous section. However, the variance of a single light curve is a poor measurement of the intrinsic variance of the variable process as \cite{Allevato13} showed. We can nevertheless use the \swift/BAT light curves of the AGN in our sample to test the PSD analysis results as we describe below.

It is important that we compute the variance using light curves with the same duration in the rest frame of each source. The highest redshift in our sample is $z=0.105$. The observed 157 month light curve of this source is 142 months long in its rest frame. We therefore removed points from all light curves until that their rest-frame duration is 142 months (we removed points from the last part of the light curves), and we computed the normalized, excess variance, \snxv,  using eq. (1) in \cite{Allevato13}. We note that, since the PSD slope is  $-1$ (or even flatter), we do need to correct the variance estimates for bias (see Section 5 in \cite{Allevato13}). The normalized excess variance measurements for each source are listed in the ninth column of Table \ref{table:sampledata}.

\begin{figure}[h]
  \centering
  \includegraphics[width=9cm]{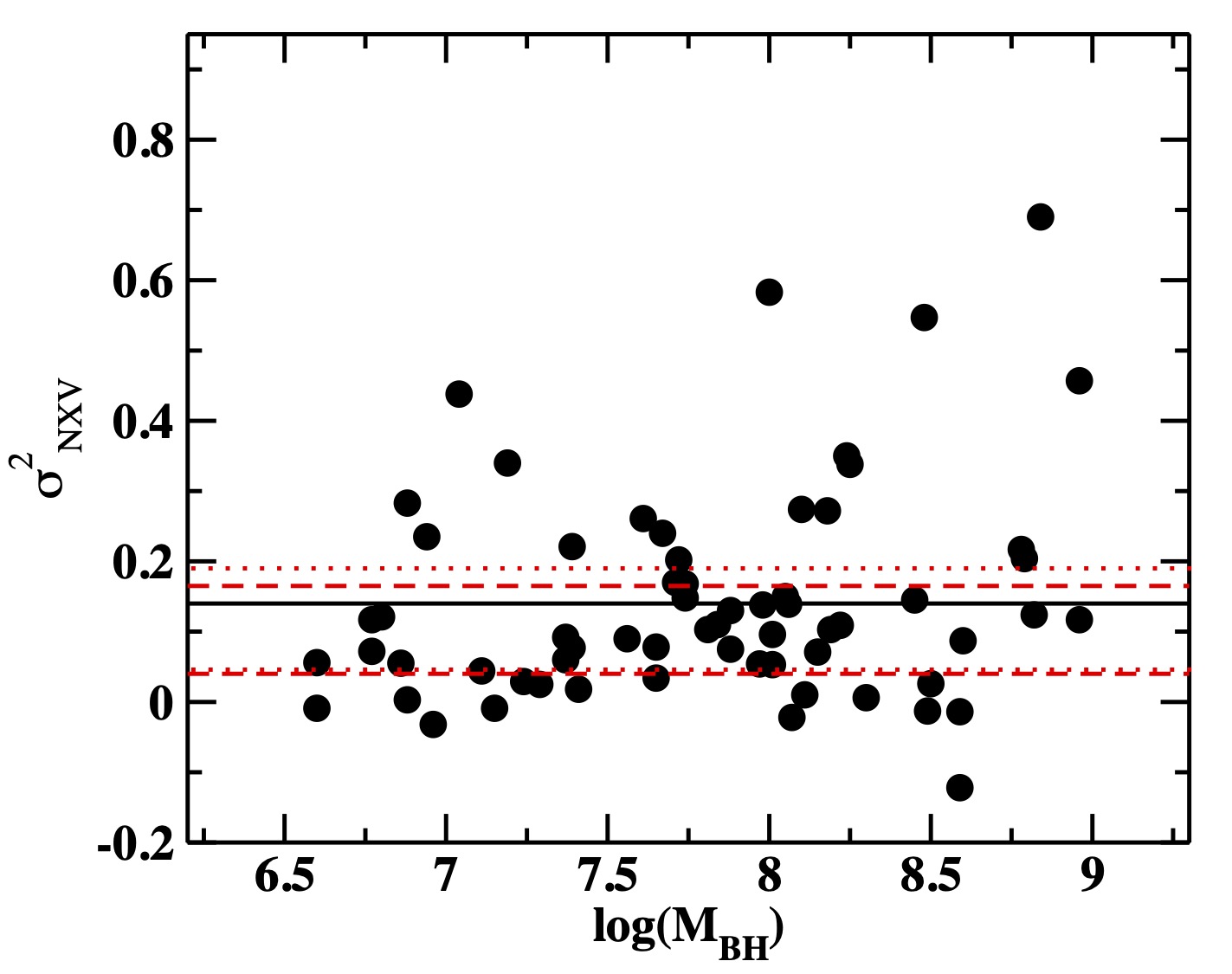}
     \caption{Plot of \snxv\ versus \lmbh, for the AGN in the various boxes plotted in Fig.\,\ref{fig:lxvsbhmsnr2}. The solid black line indicates the mean excess variance, while the dashed-dotted and the dashed lines indicate the expected variance in the case when the PSD in all AGN has a slope of $-1$ and $-0.75$, respectively.}  
        \label{fig:excessvarianceres}
\end{figure}

Figure \ref{fig:excessvarianceres} shows a plot of \snxv\ as a function of \lmbh\ for all AGN plotted within the boxes drawn in Fig.\,~\ref{fig:lxvsbhmsnr2}. These are the objects we used to study the PSD amplitude, therefore we can test the PSD results we mentioned above using the \snxv\ measurements plotted in this figure. First we investigated whether the excess variance depends on \mbh, \lx, and/or \ledd, by fitting the data with the lines: \snxv=$a_{\rm BH}+b_{\rm BH}[\log($M$_{\rm BH})-7.5]$, \snxv=$a_{\rm Lx}+b_{\rm Lx}[\log($L$_{\rm X})-43.5]$, and \snxv=$a_{\lambda_{\rm Edd}}+b_{\lambda_{\rm Edd}}[\log($\ledd$+1]$. All the best-fit $b_{\rm BH}, b_{\rm Lx}$, and $b_{\lambda_{\rm Edd}}$ values are consistent with zero, which indicates that \snxv\, does not depend om \mbh, \lx, or \ledd. This is what we would expect if the PSD is the same in all AGN, irrespective of their BH mass, luminosity and/or accretion rate. 

To test even further this hypothesis, we computed the mean excess variance of all AGN plotted in Fig.\,\ref{fig:excessvarianceres}.  We find \asnxv=0.14$\pm0.02$ (indicated by the the horizontal solid line in the same figure). If the excess variance is the same in all AGN, then \asnxv\, should be an unbiased estimate and its error should be a reliable estimate of the true error of the mean according to \cite{Allevato13} (see point 8 in their Section 8).  Using eq. (4) in \cite{Allevato13}, and the mean $PSD_{amp}$ and $PSD_{slope}$ we reported in the previous sections, we can predict the excess variance of light curves with T$_{\rm max}=142$ days, and T$_{\rm min}=2\Delta t=(2$months)/(1+$\bar{z}$), where $\bar{z}=0.023$ is the mean redshift of all AGN in the sample. The dashed lines in Fig.\,\ref{fig:excessvarianceres} indicate the expected excess variance in the case when we consider the LF PSD results (i.e. $PSD_{slope}\sim -1$ and log$(PSD_{amp})\sim 0.15$), while the dotted lines in the same figure indicate the expected excess variance in the case when we consider the FR PSD results ($PSD_{slope}\sim -0.75$ and log$(PSD_{amp})\sim 0.01$), taking into account the error on both log($PSD_{amp})$ and $PSD_{slope}$. The model excess variance in both the LF and the FR cases are fully consistent with the observed \asnxv. 

We note that the results from the fit of a line to the FR $PSD_{slope}$ vs \ledd\, data in \S\ref{sec:psdsloperes} suggested that $PSD_{slope}$ may depend on \ledd. If indeed $PSD_{slope}\propto 0.2$[log(\ledd)+1], as we found, then we would expect the AGN excess variance to increase with \ledd\, and $b_{ \lambda_{\rm Edd},FR}$= 0.05.  We find $b_{\lambda_{\rm Edd},FR}=-0.06\pm0.035$, i.e. the opposite trend. The difference between the predicted and the observed $b_{\lambda_{\rm Edd},FR}$ is at the 3$\sigma$ level. We conclude that it is rather unlikely that the intrinsic PSD slope depends on \ledd. On the other hand, in the last paragraph of the previous section we noticed a trend of decreasing PSD amplitude with increasing \ledd. Assuming the LF best-fit results, we would predict that \snxv\, should also decrease with increasing \ledd, with $b_{\lambda_{\rm Edd},SR}=-0.06$, exactly as observed: $b_{\lambda_{\rm Edd},obs}=-0.06\pm0.04$. This result further suggests that the PSD amplitude indeed decreases with increasing accretion rate.

\section{Summary and Discussion}
\label{sec:discussion}

We have used 157 month \swift/BAT light curves to compute the mean X-ray PSD of the brightest AGN in the BASS survey, in various BH mass and X--ray luminosity bins, which span two orders of magnitude in BH mass ($6.5\lesssim$\lmbh$\lesssim8.5$) and X--ray luminosity ($42.5\lesssim$log(\lx)$\lesssim44.5$; see Tables \ref{table:psdsloperes} and \ref{table:psdampres}). As we already mentioned, the AGN in our sample should be representative of the radio-quiet AGN population in the local Universe. We fitted the average power spectra with a power-law model of the form: $PSD(\nu)=PSD_{amp}(\nu/\nu_0)^{PSD_{slope}}$ (where $\nu_0=0.01$ month$^{-1}$). Our results are summarized below: 

1) The mean X--ray PSD of AGN (in the 4-195 keV band) at low frequencies ($\sim 6.5\times 10^{-3} - 0.5$ month$^{-1}$) is consistent with the power-law model. We do not detect periodicities or QPO-like features, neither do we detect any ``bending" frequencies within the sampled frequency range, where the PSD slope would steepen (at higher frequencies) or flatten (at lower frequencies).

2) The X--ray power spectrum of Seyfert 1 and Seyfert II galaxies are consistent with each other (within errors). This result shows  that the X--ray variability mechanism is consistent with being the same in both classes of AGN. This is in agreement with the AGN unification model.

3) The mean log($PSD_{amp}$) (which is defined as the PSD value at $\nu_0=0.01$/month) is  $0.15\pm 0.06$($0.01\pm 0.07$) month$^{-1}$, based on the LF(FR) PSD best-fit results. We expect a distribution of log($PSD_{amp}$) values in AGN with a given BH mass and X--ray luminosity. Our results show that the mean of the logarithm of the PSD amplitude distributions should be $\sim 0.015$month$^{-1}$, the same in all AGN, irrespective of their \mbh\, and/or \lx. 

4) We find significant evidence that the PSD amplitude of the individual AGN should decrease with increasing accretion rate as follows: $PSD_{amp}(\lambda_{\rm Edd})\propto (\lambda_{\rm Edd}/0.1)^{-0.3}$. This result implies that the variability amplitude in AGN should decrease with increasing accretion rate.

5) Just like with the PSD amplitude, our results indicate that the mean PSD slope should also be the same in all AGN, irrespective of their BH mass or luminosity. The average PSD slope is  $-0.98\pm 0.06$, when fitting only the LF PSD, and becomes $-0.74\pm 0.04$ when we fit the full range PSDs. We believe that the slope flattening when we consider high frequencies as well may be due (mainly) to aliasing effects. We also detect a significant dependence of the PSD slope on $\lambda_{\rm Edd}$, when we consider the full range results. The power spectrum slope appears to flatten with increasing $\lambda_{\rm Edd}$. As we discuss below, this result is also due to aliasing effects, and does not imply an intrinsic dependence of the the PSD slope on accretion rate.

6) The normalized excess variance, \snxv, does not depend on BH mass or \lx. This result is consistent with the PSD results. If the average X--ray PSD is the same in all AGN, then its integral, i.e. \snxv, should be independent of the AGN parameters, as we observe. The mean \snxv\ is consistent with the expected excess variance either when we assume a PSD slope of $-1$ or $-0.75$. 

\subsection{The universal shape of the mean X--ray PSD of AGN}
\label{sec:universalpsd}

\begin{figure}[h]
  \centering
  \includegraphics[width=9cm]{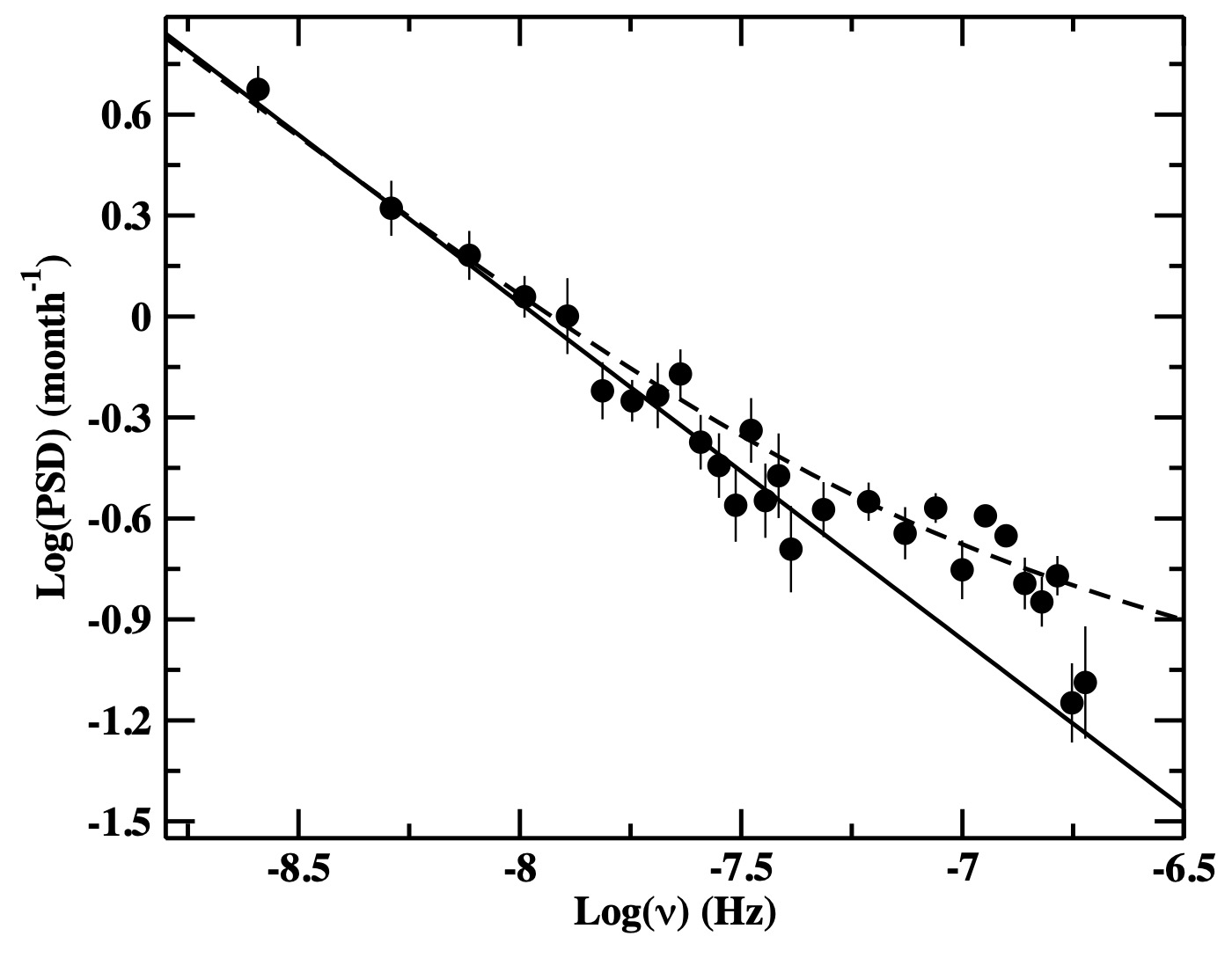}
     \caption{The mean PSD of the variable AGN in our sample. The solid line shows the best-fit power-law model to the data up to $10^{-7.4}$ Hz, and the dashed line shows the same model with aliasing effects included (see \S\ref{sec:universalpsd} for details).}  
        \label{fig:meanpsd}
\end{figure}

Our results suggest that, on average, the X--ray PSD (in the 14-195 keV band) should be the same in all AGN. If that is the case, then we can compute the mean AGN PSD if we use the periodograms of all the variable sources in our sample. To this end, we computed the mean PSD of all variable AGN in our sample (corrected from the Poisson noise) as explained in \S\ref{sec:psdestimation}. The resulting mean PSD is shown in Fig.\,\ref{fig:meanpsd}. We have smoothed the high frequency PSD points (above 0.1 month$^{-1}$) as explained in the same section. 

The mean PSD is fully consistent with a simple power-law model. The solid line in this figure shows the best-fit model when we fit the data below $10^{-7.4}$ Hz (=0.1/month) with a power-law model defined by eq.\,\ref{eq:linemodel}. This line provides a very good fit to the LF part of the PSD ($\chi^2=10.6/13$dof) with a best fit slope of $-1.00\pm 0.06$. The PSD amplitude is larger than the mean amplitude we report in \S\ref{sec:psdampresults}, but that is because here we consider only the variable AGN (in order to determine the PSD slope correctly). 

The fact that a $-1$ power-law model fits so well the LF part of the mean PSD strongly supports the hypothesis that there is no PSD flattening even at frequencies as low as $\sim 3\times 10^{-9}$ Hz. For example, if there was a low-frequency flattening for low mass BHs in our sample, we would expect the lowest-frequency points in the mean PSD to show a much larger scatter, when compared to the higher frequency points, but that is not the case. We believe that our results clearly show that the average AGN PSD follows a power-law shape with a slope of $-1$ down to frequencies as low as  a few $10^{-9}$ Hz (this frequency corresponds to a time scale of $\sim 12$ years).  

The mean PSD flattens at frequencies higher than $\sim 4\times 10^{-8}$ Hz. We do not believe that this is due to a systematic bias in the computation of the Poisson noise level (see Fig.\,\ref{fig:pnoiseratio} and the discussion in \S\ref{sec:noiseestimation}). As we discussed in \S\ref{sec:psdfit}, the high frequency flattening could be due to aliasing effects. The dashed line in Fig.\,\ref{fig:meanpsd} shows the best-fit power-law model (the one plotted with the solid line) when we also include the aliasing effects in the case when the PL extends to $2\times 10{-6}$ Hz, and the light curve is evenly sampled. The frequency of $2\times 10{-6}$ Hz indicates the high frequency break in the 2--10 keV PSD of an AGN with \lmbh=7.5, according to the model defined by eq. (4) in \cite{Gonzalez-Martin2012}. The agreement between the dashed line and the mean PSD at high frequencies is quite good. This result supports the fact that the PSD flattening we detect at high frequencies is probably due to aliasing effects.

Support to the hypothesis of aliasing affecting the measured PSD slope when we consider the full frequency range is also provided by the apparent dependence of the PSD slope on $\lambda_{\rm Edd}$, when we consider the FR results (see \S \ref{sec:psdsloperes}). According to \cite{McHardy06}, the PSD bending frequency should increase with increasing $\lambda_{\rm Edd}$. In this case, we would also expect the aliasing effects to become stronger, hence resulting into a flattening of the PSD slope with increasing $\lambda_{\rm Edd}$, as observed.

The hard X--ray PSD should steepen above a bending frequency, $\nu_b$, just like in the 2--10 keV band. The fact that such a bending must exist, and is probably located at a frequency similar to the high-frequency break in the 2-10 keV PSDs is supported by the agreement of the dashed line and the mean PSD in Fig.\,\ref{fig:meanpsd}. If the intrinsic hard X--ray PSD would extend to very high frequencies without bending to a slope steeper than $-1$ then, irrespective of the sampling pattern of the \swift/BAT light curves, the aliasing effects would be much stronger, and the high frequency flattening in Fig.\,\ref{fig:meanpsd} would be much more pronounced. If the bending frequency does not vary significantly with energy, we would then expect $\nu_b\sim470-4.7$ month$^{-1}$ (i.e $\sim 1.8\times 10^{-4}-1.8\times 10^{-6}$ Hz) for AGN with \lmbh=6.5-8.5 (which is the majority of the AGN in our sample). These frequencies cannot be sampled with the current \swift/BAT light curves. {\it NuSTAR} light curves would be necessary in order to detect and investigate the bending frequency properties at hard X-rays. 

If the bending frequency does not depend significantly on energy, this would imply that the hard X--ray PSD in AGN extends with a slope of $-1$ over $\sim 2.8 - 4.8$ decades 
for AGN with \lmbh=8.5-6.5, respectively. This is a large frequency range and, at least for AGN with BH masses smaller than a few 10$^7$M$_{\odot}$, this result would have interesting implications. For example, the Ark 564 PSD shows a significant low-frequency break to slope zero at $\sim 10^{-6}$ Hz, $\sim 3$ orders of magnitude below the high-frequency break \citep{McHardy07,Papadakis02}. This is the only AGN where such a low frequency flattening has been detected in its X--ray PSD. It is quite possible then that such PSDs are not very common in AGN. As we argued above, if this was common, we would expect the low frequency points in the PSD plotted in Fig.\,\ref{fig:meanpsd} to show evidence of a low frequency flattening and to be very noisy.

\subsection{Comparison with past results}

Table 3 in SM13 lists their best-fit log($PSD_{amp}$) and $PSD_{slope}$ results when they fit an unbroken power-law model to the data. We fit the same model to the \swift/BAT power spectra. According to the SM13 results, the PSD at $\nu_0=0.01$ month$^{-1}$ should be given by $PSD_{SM13}(\nu_0)=A_{SM13}(\nu_0/\nu_{SM13})^{-PSDslope,SM13}$, where $\nu_{SM13}=10^{-6}$ Hz, and it should be consistent with the amplitude of our PSD model. Using the SM13 best-fit results, we found that $\overline{\log[PSD_{SM13}(\nu_0)]}=-0.13\pm 0.10$, which is entirely consistent with our LF, mean log$(PSD_{amp})$ measurement. 

Furthermore, SM13 find a mean PSD slope of $-0.78\pm 0.29$, which becomes $-0.92\pm 0.22$ when they restrict the average to only the 16  best-fit slopes that were fully constrained (out of the 30 sources in their sample). The SM13 results are very similar to ours. We find a mean slope of $\sim -0.75$ when we fit the full-range PSDs, and a steeper slope of $\sim -1$ when we consider only the LF part of the power spectrum. As we mentioned above, we believe that the $-1$ value is probably more representative of the intrinsic PSD slope, mainly because aliasing effects may flatten the PSD at high frequencies. Such effects should not be important in the SM13 power spectra because they used light curves with a bin size of 1 day. Since the power at frequencies higher than $\sim 1/$day should be significantly smaller than the power at low frequencies, aliasing effects should not be important in this case. On the other hand, as SM13 comment, their best-fit PSD slopes are well constrained only when the power spectra are well above the Poisson noise. The fact that the mean slope in this case is closer to our mean in the case of the LF best-fit results reinforces our belief that $-1$ is probably closer to the intrinsic PSD slope of the AGN X--ray PSD at low frequencies. 

\subsection{Comparison with the 2-10 keV band PSD results}

As we have already mentioned in the Introduction, many works in the past have shown that the low frequency PSD of AGN at energies below 10 keV has a slope of $-1$. As we discussed in \S\ref{sec:universalpsd}, this is probably the case of the low frequency, hard X--ray PSD in AGN as well. Therefore, our results suggest that the slope of the low-frequency PSD in the AGN does not depend on energy. This is opposite to what is observed at high frequencies in the 2--10 keV band PSD, where the slope above the high frequency break frequency appears to flatten with energy.

Our results are in agreement with the results of \cite{Paolillo23}, who find that the 2--10 keV PSD has the same shape in all AGN, over a large range of BH mass and luminosity. Both in the 2--10 and in the 14-195 keV bands the universal PSD slope at low frequencies is consistent with $-1$. On the other hand, \cite{Paolillo23} find that $PSD_{2-10 keV}(\nu_b)\times\nu_b=0.008\pm0.002$, which is smaller than our estimate of $0.014\pm 0.003$. This could imply that the PSD amplitude is larger at higher energies, although the difference is not statistically significant.  

Excess variance studies in the 2--10 keV band indicate that $PSD(\nu_b)\times \nu_b\propto$\ledd$^{-0.8}$ \citep{Ponti12,Paolillo17}. If this holds in the 14-195 keV band as well, it would imply a similar dependence of $PSD_{amp}$ on \ledd. We do detect a significant anti-correlation between PSD amplitude and \ledd\ in our data (see \S\ref{sec:psdampresults}). We find that $PSD_{amp}\propto$(\ledd)$^{-0.30\pm0.12}$, which is flatter than the anti-correlation  which the 2--10 keV data imply (although the 2--10 keV relation is not well defined, with no error given on the slope). Perhaps the PSD amplitude--\ledd\ anti-correlation at energies higher than 15 keV is not as steep as the one at lower energies, although a more detailed study of this effect in the 2-10 keV band is necessary to understand if there are differences between the two bands.  

\subsection{Comparison with the GBH PSD}

Both AGN and GBHs are power by accretion of matter on BHs, a major difference being the mass of the BH at their center. It has long been suggested that the X--ray variability mechanism is the same in both systems. However, the PSD of GBHs depends on the “state” of the system. There are two main states, the hard and the soft, in GBHs. The PSD in the soft state is usually well described by a single bending power-law, with a slope equal/steeper than $-2$ above $\nu_b$, and a slope of $-1$ extending to very low frequencies. In the hard state, the PSD is more complex, and is usually modeled as a mixture of zero-centred Lorentzian components. It is true that similar power spectra from the hardest part of the hard state show similar shape to the soft state PSDs \citep[e.g.][]{Heil15}, however, in general, black hole X-ray binary hard states show power spectra with slopes $\sim -1$ over at least 1-2 decades in frequency, which bend to a constant, i.e. to a slope of zero, at lower frequencies \cite[e.g.][]{Remillard06}. Based on the shape of the AGN PSDs (e.g. \cite{McHardy04, Markowitz03, Uttley05}; see also \cite{Paolillo23}), as well as the scaling of $-1$ to $-2$ bending frequencies between AGN and GBHs \citep{McHardy06}, it has been argued that almost all of the X--ray bright AGN should be the analogue of the GBHs in their soft state. SM13 reached the same conclusion because they did not find any evidence for a low frequency break to a slope of zero in the AGN \swift/BAT PSDs they studied.

\begin{figure}[h]
  \centering
  \includegraphics[width=9cm]{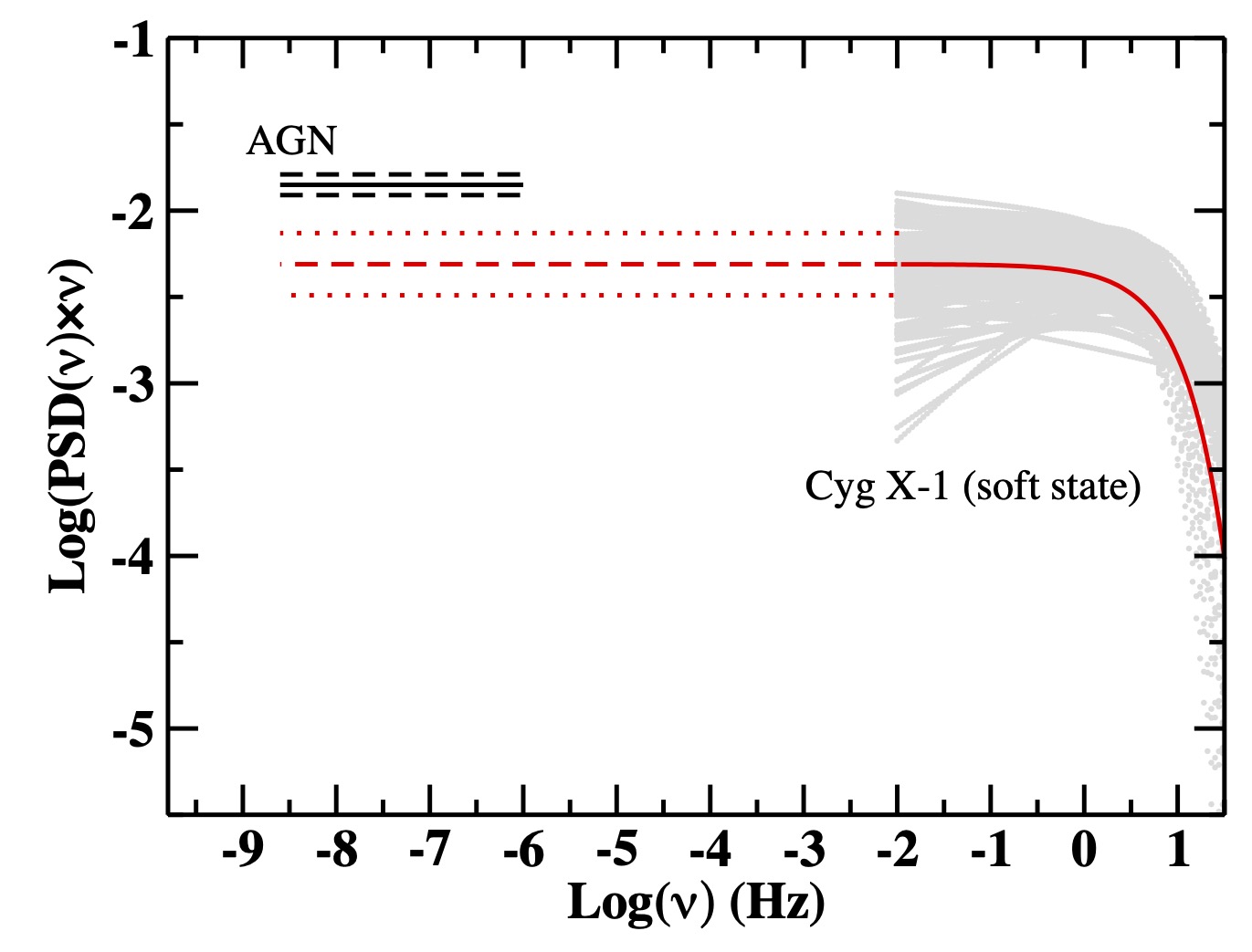}
     \caption{The grey lines show the Cyg X--1 PSDs in its canonical soft state according to \cite{Axelsson06}. The solid red line indicates their mean, while the red dashed and dotted lines show its extension, and 1$\sigma$ scatter, respectively, to lower frequencies. The solid black and brown lines indicate the mean AGN PSD when we consider our LF and FR best-fit results, respectively (dashed lines indicate the 1$\sigma$ scatter due to the $PSD_{amp}$ error). }  
        \label{fig:agncygx1psd}
\end{figure}

We compared our results with results from the PSD analysis of Cyg X-1 light curves when the source is in the soft state. Such a comparison between the timing properties of AGN and Cyg X-1 in its soft state has been done many times in the past \citep[e.g.][]{Uttley02, McHardy06}. The solid, grey lines in Fig.\,\ref{fig:agncygx1psd} show the Cyg X-1 PSDs (in $PSD(\nu)\times \nu$ representation), when the source is in its soft state. The data are taken from \cite{Axelsson06}, who analyzed all the the archival Cyg X--1 data from the Rossi X-ray Timing Explorer (RXTE) from 1996 to 2003 in the 2--9 keV band, and in the frequency range of 0.01–25 Hz. They used various models to fit the PSDs, and one of them was a cut-off power-law (their eq. 2). The grey lines show the best-fit models, when the PSDs were fitted only by the cut-off power law. According to the authors this model fits well the PSD only when the source is in its canonical soft state. Figure \ref{fig:agncygx1psd} shows that the amplitude of the Cyg X-1 PSD in the soft state varies by a factor $\sim 15$ or so. The PSD amplitude probably varies among AGN as well, as we have discussed above, but what is interesting is to compare the mean AGN PSD with the mean Cyg X-1 PSD in its soft state.

The solid red line shows the mean Cyg X-1 PSD. The dashed line shows its extension to lower frequencies, while the dotted lines indicate the 1$\sigma$ spread of the PSD normalization around its mean. The solid black line shows the mean best-fit PSD of the AGN our sample. The frequency range extends to $10^{-6}$ Hz, as SM13 showed the the AGN PSDs extend as a power-law up to that frequency. The dashed lines indicates the $1\sigma$ uncertainty of the best-fit to the mean $PSD_{amp}$. 
Clearly, the mean AGN PSD agrees very well with the Cyg X-1 PSD in its soft state. The difference in BH mass between AGN in our sample and Cyg X-1 is $\sim 10^6-10^8$, while the difference in frequency between the Cyg X-1 and AGN PSD is over 4--5 orders of magnitude. Despite these differences the PSD slopes are identical, and the mean PSD amplitude differs only by a factor of 3, maximum. We suspect that this difference is probably due to the difference in energy bands (2--9 KeV for Cyg X-1 as opposed to 14-195 keV for  AGN). We plan to investigate this issue further in the near future. 

We note, that the agreement between the mean AGN PSD and the Cyg X-1 PSD in its soft state is quite plausible but, strictly speaking, suggestive at this point. For example, hard state low-frequency breaks can be seen from 0.01 Hz and higher, which would be 10$^{-9}$ Hz or higher for a 10$^8$M$_{\odot}$ mass black hole, assuming linear mass-scaling. As we discussed in \S\ref{sec:universalpsd}, if that were the case we would have detected it in the mean AGN PSD. However, we cannot be certain about the location of the high frequency breaks in these objects (at the hard X-ray PSD). If the mean AGN PSD were to be similar to the GBHs PSD in the hard state, then the $-1$ to $-2$ breaks in AGN at energies higher than 15 keV must be located at frequencies much lower than the ones we have detected at energies below 10 keV.

If, on the other hand, the mean AGN PSD is like the mean Cyg X-1 in its soft state, then it is possible that the physical properties are similar in both systems. It is generally believed that the accretion disc extends to its innermost radius, and the X-ray corona is less luminous than the disc in GBHs when in their soft state. This could be then the case in AGN as well. The disc should extend to its innermost radius and its emission would dominate the output in these objects as well (as is indeed the case with the ``big blue bump" that is observed in these objects). Given the inner disc radius, the corona cannot be located within the disc, and it has to be located on top of it (or above the central BH), illuminating the disc and causing all the X--ray reflection features (i.e. soft excess, broad iron line and Compton hump) as well as the X--ray reverberation phenomena (like e.g. driving the UV/optical variability and producing the delays between X--rays and UV/optical variations). 

\begin{acknowledgements}
We would like to thank the referee for their comments which helped us improve the original version of the paper.
\end{acknowledgements}

\bibliographystyle{aa}
\bibliography{swift-bat-psds}

\clearpage

\onecolumn

\begin{appendix}

\section{}

\begin{longtable}{lcccccccc}
\caption{\label{table:sampledata} AGN in the sample and their properties.}\\
\hline\hline
Source & DR2 & \fobs\tablefootmark{(b)}  & $z$\tablefootmark{(a)} & log(\mbh)\tablefootmark{(a)} & log(L$_{\rm X}$)\tablefootmark{(c)} & $\chi^2$/dof\tablefootmark{(d)} & V/NV\tablefootmark{(e)} & $\sigma^{2}_{NXV}$\tablefootmark{(f)} \\
 & Type\tablefootmark{(a)} & & & & (erg/s) & & & \\
\hline
\endfirsthead
CenA& Sy2 & 1346.34 & 0.0019 & 7.77 & 42.33 & 80640.65/156 & V & 0.165\\
NGC4151& Sy1 & 618.88 & 0.0032 & 7.56 & 43.30 & 25562.30/156 & V & 0.092\\
NGC2110& Sy2 & 328.93 & 0.0075 & 8.78 & 43.60 & 7188.11/156 & V & 0.221\\
NGC4945& Sy2 & 282.11 & 0.0023 & 6.15 & 42.33 & 2129.02/156 & V & 0.112\\
NGC4388& Sy2 & 278.91 & 0.0083 & 6.94 & 43.05 & 5008.13/155 & V & 0.237\\
CircinusGalaxy& Sy2 & 273.17 & 0.0015 & 6.23 & 42.06 & 525.07/156 & V & 0.010\\
IC4329A& Sy1 & 263.25 & 0.0160 & 7.65 & 44.16 & 1371.12/156 & V & 0.078\\
NGC5506& Sy1.9 & 239.40 & 0.0060 & 7.24 & 43.24 & 597.65/153 & V & 0.029\\
4C+50.55& Sy1.9 & 210.38 & 0.0154 & 7.97 & 44.01 & 1405.12/156 & V & 0.054\\
MCG-5-23-16& Sy1.9 & 209.62 & 0.0084 & 7.65 & 43.44 & 827.79/156 & V & 0.035\\
NGC4507& Sy2 & 184.54 & 0.0117 & 7.81 & 43.90 & 993.72/156 & V & 0.103\\
NGC3783& Sy1 & 173.84 & 0.0090 & 7.37 & 43.45 & 736.34/156 & V & 0.060\\
NGC7172& Sy2 & 160.02 & 0.0085 & 8.15 & 43.18 & 875.63/152 & V & 0.074\\
Mrk3& Sy1.9 & 150.12 & 0.0138 & 8.96 & 44.02 & 1343.69/156 & V & 0.115\\
CygnusA& Sy2 & 145.23 & 0.0565 & 9.43 & 44.97 & 947.41/156 & V & 0.080\\
Mrk348& Sy1.9 & 144.81 & 0.0147 & 6.8 & 43.83 & 1357.28/156 & V & 0.122\\
UGC3374& Sy1 & 141.21 & 0.0202 & 6.61 & 44.03 & 1260.05/153 & V & 0.212\\
IRAS05078+1626& Sy1 & 119.80 & 0.0174 & 6.88 & 43.74 & 994.33/151 & V & 0.286\\
1RXSJ173728.0-290759& Sy1.9 & 119.04 & 0.0218 & 7.84 & 44.05 & 719.70/145 & V & 0.112\\
3C111& Sy1 & 118.67 & 0.0497 & 8.45 & 44.79 & 692.25/152 & V & 0.144\\
NGC3227& Sy1 & 112.47 & 0.0033 & 6.77 & 42.77 & 648.91/153 & V & 0.071\\
NGC3516& Sy1 & 112.42 & 0.0087 & 7.39 & 43.26 & 1561.91/156 & V & 0.224\\
4U1344-60& Sy1.9 & 111.79 & 0.0128 & 9.09 & 43.52 & 211.47/156 & V & -0.002\\
Mrk926& Sy1 & 110.22 & 0.0477 & 7.98 & 44.72 & 605.22/155 & V & 0.139\\
ESO103-35& Sy1.9 & 106.86 & 0.0135 & 7.37 & 43.64 & 553.05/156 & V & 0.093\\
NGC5252& Sy2 & 103.15 & 0.0230 & 9.0 & 44.05 & 1352.73/155 & V & 0.426\\
3C390.3& Sy1 & 102.87 & 0.0556 & 8.64 & 44.82 & 772.56/156 & V & 0.086\\
Mrk509& Sy1 & 100.14 & 0.0347 & 8.05 & 44.37 & 526.51/153 & V & 0.150\\
MR2251-178& Sy1 & 99.53 & 0.0645 & 8.19 & 44.95 & 370.15/153 & V & 0.103\\
NGC6300& Sy2 & 96.37 & 0.0031 & 6.77 & 42.29 & 701.76/156 & V & 0.117\\
3C120& Sy1 & 95.38 & 0.0331 & 7.74 & 44.32 & 697.27/155 & V & 0.179\\
ESO506-27& Sy2 & 90.68 & 0.0242 & 8.84 & 44.17 & 1038.69/154 & V & 0.672\\
NGC4593& Sy1 & 88.30 & 0.0083 & 6.88 & 43.10 & 233.36/153 & V & 0.004\\
NGC5548& Sy1 & 86.47 & 0.0167 & 7.72 & 43.62 & 1129.77/156 & V & 0.195\\
NGC5728& Sy1.9 & 84.74 & 0.0103 & 8.25 & 43.24 & 611.12/154 & V & 0.341\\
NGC1275& Sy1.9 & 82.64 & 0.0168 & 8.6 & 43.65 & 649.27/156 & V & 0.089\\
3C382& Sy1 & 82.33 & 0.0579 & 8.01 & 44.72 & 473.91/156 & V & 0.097\\
NGC7582& Sy2 & 82.28 & 0.0052 & 7.74 & 43.24 & 523.49/155 & V & 0.148\\
NGC3081& Sy2 & 81.89 & 0.0081 & 7.67 & 43.18 & 526.76/156 & V & 0.243\\
NGC3281& Sy2 & 81.21 & 0.0111 & 8.22 & 43.46 & 479.64/156 & V & 0.108\\
NGC788& Sy2 & 77.51 & 0.0137 & 8.18 & 43.61 & 585.38/156 & V & 0.276\\
NGC1142& Sy2 & 74.31 & 0.0287 & 8.96 & 44.31 & 730.35/156 & V & 0.451\\
Ark120& Sy1 & 74.29 & 0.0326 & 8.07 & 44.16 & 176.15/156 & NV & -0.022\\
NGC526A& Sy1.9 & 73.91 & 0.0189 & 8.06 & 43.63 & 699.07/156 & V & 0.140\\
NGC7469& Sy1 & 70.63 & 0.0160 & 6.96 & 43.51 & 182.87/155 & NV & -0.033\\
IC5063& Sy2 & 67.76 & 0.0113 & 8.24 & 43.29 & 545.68/156 & V & 0.353\\
ESO297-18& Sy2 & 67.37 & 0.0252 & 8.5 & 44.10 & 219.67/156 & V & 0.025\\
NGC6814& Sy1 & 64.47 & 0.0058 & 7.04 & 42.62 & 777.90/150 & V & 0.439\\
NGC1365& Sy1 & 63.52 & 0.0051 & 6.6 & 42.39 & 322.18/156 & V & 0.057\\
IRAS05589+2828& Sy1 & 62.28 & 0.0331 & 8.0 & 44.14 & 836.57/147 & V & 0.578\\
Mrk1040& Sy1 & 62.22 & 0.0163 & 7.41 & 43.51 & 193.46/154 & NV & 0.019\\
Mrk110& Sy1 & 60.95 & 0.0352 & 7.29 & 44.16 & 214.51/156 & V & 0.024\\
MCG-6-30-15& Sy1 & 59.53 & 0.0071 & 6.6 & 42.78 & 174.65/155 & NV & -0.010\\
ESO141-55& Sy1 & 58.77 & 0.0371 & 7.99 & 44.18 & 174.08/156 & NV & -0.028\\
LEDA178130& Sy2 & 58.27 & 0.0352 & 7.88 & 44.17 & 311.22/156 & V & 0.128\\
NGC7314& Sy1.9 & 57.42 & 0.0046 & 6.3 & 42.16 & 211.06/153 & V & 0.035\\
Mrk1210& Sy1.9 & 57.40 & 0.0136 & 6.86 & 43.35 & 198.27/155 & NV & 0.055\\
Mrk6& Sy1 & 56.70 & 0.0190 & 8.1 & 43.62 & 586.66/156 & V & 0.274\\
LEDA138501& Sy1 & 55.63 & 0.0497 & 8.01 & 44.42 & 234.89/156 & V & 0.051\\
NGC4992& Sy2 & 54.42 & 0.0252 & 8.59 & 43.93 & 195.39/155 & NV & -0.012\\
LEDA168563& Sy1 & 54.10 & 0.0283 & 7.88 & 43.98 & 273.67/156 & V & 0.077\\
NGC7603& Sy1 & 52.96 & 0.0287 & 8.59 & 43.92 & 129.26/150 & NV & -0.119\\
NGC235A& Sy1.9 & 52.14 & 0.0221 & 8.49 & 43.71 & 180.41/156 & NV & -0.001\\
NGC6860& Sy1.9 & 51.52 & 0.0147 & 7.71 & 43.33 & 338.31/156 & V & 0.167\\
4C+74.26& Sy1 & 51.18 & 0.1050 & 9.83 & 45.10 & 183.93/156 & NV & 0.025\\
2MASSJ07594181-3843560& Sy1 & 51.15 & 0.0400 & 8.82 & 44.21 & 180.72/156 & NV & 0.125\\
LEDA38038& Sy2 & 49.46 & 0.0277 & 8.11 & 43.86 & 184.09/156 & NV & 0.015\\
LEDA86269& Sy2 & 49.38 & 0.0105 & 7.98 & 43.02 & 484.86/152 & V & 0.496\\
NGC612& Sy2 & 48.97 & 0.0299 & 8.28 & 44.26 & 322.18/156 & V & 0.210\\
ESO362-18& Sy1 & 48.89 & 0.0125 & 7.11 & 43.17 & 216.20/156 & V & 0.046\\
NGC973& Sy2 & 48.41 & 0.0157 & 8.48 & 43.18 & 1045.03/156 & V & 0.540\\
Fairall9& Sy1 & 48.39 & 0.0459 & 8.3 & 44.34 & 172.59/156 & NV & 0.007\\
HE1143-1810& Sy1 & 48.28 & 0.0326 & 7.39 & 44.05 & 215.05/156 & V & 0.074\\
MCG-1-24-12& Sy2 & 46.38 & 0.0196 & 7.66 & 43.50 & 213.07/156 & V & 0.157\\
LEDA166252& Sy1 & 44.90 & 0.0249 & 8.79 & 43.76 & 302.68/156 & V & 0.204\\
1RXSJ165605.6-520345& Sy1 & 43.23 & 0.0538 & 8.33 & 44.44 & 177.28/156 & NV & 0.142\\
Mrk1498& Sy1 & 42.97 & 0.0558 & 7.19 & 44.48 & 493.22/156 & V & 0.321\\
Mrk79& Sy1 & 42.72 & 0.0221 & 7.61 & 43.64 & 457.84/156 & V & 0.265\\
NGC4051& Sy1 & 42.49 & 0.0020 & 6.13 & 41.69 & 169.25/156 & NV & 0.002\\
2MASXiJ1802473-145454& Sy1 & 42.36 & 0.0343 & 7.98 & 43.97 & 157.24/152 & NV & 0.017\\
1RXSJ174155.3-121157& Sy1 & 41.54 & 0.0376 & 8.03 & 44.13 & 168.96/154 & NV & 0.042\\
ESO548-81& Sy1 & 41.21 & 0.0144 & 7.96 & 43.26 & 337.36/156 & V & 0.253\\
3C445& Sy1 & 39.82 & 0.0561 & 7.89 & 44.56 & 150.83/153 & NV & -0.016\\
ESO490-26& Sy1 & 39.49 & 0.0252 & 7.15 & 43.67 & 224.91/155 & V & 0.087\\
IGRJ21277+5656& Sy1 & 39.43 & 0.0149 & 7.15 & 43.21 & 172.84/156 & NV & -0.009\\
ESO511-30& Sy1 & 39.27 & 0.0229 & 7.23 & 43.62 & 202.05/155 & V & -0.014\\
Fairall272& Sy2 & 39.15 & 0.0222 & 8.11 & 43.73 & 223.64/155 & V & 0.170\\
IC4709& Sy2 & 39.08 & 0.0169 & 7.81 & 43.40 & 171.50/156 & NV & -0.096\\
NGC7213& Sy1 & 39.04 & 0.0048 & 7.13 & 42.32 & 164.71/155 & NV & 0.105\\
NGC4235& Sy1 & 38.62 & 0.0079 & 7.28 & 42.36 & 443.00/153 & V & 0.474\\
NGC1068& Sy1.9 & 37.90 & 0.0035 & 6.93 & 42.65 & 182.20/156 & NV & 0.040\\
PictorA& Sy1 & 37.32 & 0.0350 & 6.77 & 43.98 & 264.56/156 & V & 0.165\\
Mrk704& Sy1 & 36.84 & 0.0295 & 8.24 & 43.76 & 169.00/151 & NV & 0.046\\
NGC3079& Sy2 & 36.74 & 0.0035 & 6.38 & 42.70 & 203.91/156 & V & 0.030\\
Fairall51& Sy1 & 36.51 & 0.0139 & 7.11 & 43.18 & 181.93/156 & NV & 0.040\\
NGC7319& Sy2 & 36.45 & 0.0226 & 8.1 & 43.71 & 209.13/156 & V & 0.313\\
NGC1194& Sy2 & 36.22 & 0.0136 & 7.83 & 43.62 & 154.44/156 & NV & -0.053\\
CTS109& Sy1 & 35.80 & 0.0300 & 7.84 & 43.89 & 228.57/153 & V & 0.050\\
Z121-75& Sy1 & 35.66 & 0.0331 & 7.27 & 43.94 & 177.16/152 & NV & -0.011\\
LEDA170194& Sy2 & 35.33 & 0.0363 & 8.35 & 44.01 & 221.19/153 & V & 0.358\\
\hline
\end{longtable}

\begin{tablenotes}
\item{{\bf Notes.} \tablefoottext{a}{Taken from \cite{Koss2022}. They classified the unbeamed AGN in three categories, Sy1, Sy1.9 and Sy2, based on the presence of:  1) broad H$\beta$ lines, 2) narrow H$\beta$ but broad H$\alpha$, and 3) only narrow optical lines in their optical spectra, respectively.} 
\tablefoottext{b}{Taken from the Swift-BAT 157 month survey (in units of $10^{-12}$ ergs/s/cm$^2$).}
\tablefoottext{c}{Logarithm of the intrinsic X--ray luminosity in the 14-195 keV band, computed as explained in \S\ref{sec:sample}.} 
\tablefoottext{d}{The $\chi^2$, and the number of degrees of freedom (dof) when we fit a constant line to the light curves.}
\tablefoottext{e}{Classification of the variable (V) and non-variable (NV) sources, depending on $p_{null}$ (see \S\ref{sec:sample} for details).}
\tablefoottext{f}{The normalized excess variance, \snxv, computed as explained in \S\ref{sec:snxvresults}.}}
\end{tablenotes}

\end{appendix}

\end{document}